\documentclass[prd,reprint,nofootinbib,superscriptaddress,preprintnumbers,balancelastpage,longbibliography,nobibnotes,tightenlines,10pt,flushbottom]{revtex4-1}
\usepackage{graphicx,epsfig,psfrag,bm,amsmath,amssymb}
\usepackage{dcolumn}
\usepackage{bm}
\usepackage[dvipsnames]{xcolor}
\usepackage{braket}
\usepackage{mathrsfs,amsfonts,color}
\usepackage{comment}
\usepackage{soul}
\usepackage[caption=false]{subfig}
\usepackage{setspace}
\usepackage{siunitx}
\usepackage[page,toc,titletoc,title]{appendix}
\usepackage{hyperref}
\usepackage[capitalise]{cleveref}

\definecolor{linkcolor}{rgb}{0.0, 0.28, 0.67}
\hypersetup{colorlinks=true,
linkcolor=black,
citecolor=linkcolor,
urlcolor=linkcolor,
linktocpage=true
}

\newcommand{\be}{\begin{eqnarray}}
\newcommand{\ee}{\end{eqnarray}}

\newcommand{\w}{\omega}

\newcommand{\pd}{\partial}

\newcommand{\p}{\varphi}

\renewcommand{\vec}[1]{\boldsymbol{#1}}

\begin{document}

\preprint{CERN-TH-2024-151}

\title{Dielectric Haloscopes as Gravitational Wave Detectors}

\author{Valerie~Domcke}
\affiliation{Theoretical Physics Department, CERN, 1 Esplanade des Particules, CH-1211 Geneva 23, Switzerland}

\author{Sebastian~A.~R.~Ellis}
\affiliation{D\'epartement de Physique Th\'eorique, Universit\'e de Gen\`eve, 
24 quai Ernest Ansermet, 1211 Gen\`eve 4, Switzerland}

\author{Joachim~Kopp}
\affiliation{Theoretical Physics Department, CERN, 1 Esplanade des Particules, CH-1211 Geneva 23, Switzerland}
\affiliation{PRISMA+ Cluster of Excellence \& Mainz Institute for Theoretical Physics,
             55128 Mainz, Germany}

\begin{abstract}
We argue that dielectric haloscopes like MADMAX, originally designed for detecting axion dark matter, are also very promising gravitational wave detectors.
Operated in resonant mode at frequencies around $\mathcal{O}(\SI{10}{GHz})$, these detectors benefit from enhanced gravitational wave to photon conversion at the surfaces of a stack of thin dielectric disks. Since the gravitational wave is relativistic, there is an additional enhancement of the signal compared to the axion case due to increased conversion probability of gravitational waves to photons in the vacuum between the disks. A gravitational wave search using a dielectric haloscope imposes stringent requirements on the disk thickness and placement, but relaxed requirements on the disk smoothness. An advantage is the possibility of a broadband or hybrid resonant/broadband operation mode, which extends the frequency range down to $\mathcal{O}(\SI{100}{MHz})$. We show that strain sensitivities down to $\SI{e-21}{Hz^{-1/2}} \times  (\SI{10}{GHz}/f)$ will be possible in the coming years for the broadband setup, while a resonant setup optimized for gravitational waves could even reach $\SI{3e-23}{Hz^{-1/2}} \times (\SI{10}{GHz}/f)$ with current technology. 
\end{abstract}

\maketitle

\section{Introduction}

High frequency gravitational waves (GWs) are a unique window to probe rare exotic astrophysical events such as light primordial black hole mergers~\cite{Franciolini:2022htd} or light boson superradiance around primordial black holes~\cite{Brito:2015oca}, as well as cosmological processes in the very early Universe~\cite{Aggarwal:2020olq}. The challenges that have to date prevented any detection above the \SI{100}{Hz} band probed by LIGO, VIRGO, and KAGRA \cite{LIGOScientific:2014pky, VIRGO:2014yos, KAGRA:2020tym} lie in the very small GW amplitude characteristic of sources in this frequency band and the resulting need for extremely sensitive detectors~\cite{Aggarwal:2020olq}.

Many detector concepts have been put forward recently, several of them based on techniques used in axion searches~\cite{Ejlli:2019bqj, Berlin:2021txa, Domcke:2022rgu, Bringmann:2023gba, Domcke:2023bat, Navarro:2023eii, Valero:2024ncz, Domcke:2024mfu}. In this article we propose to use a dielectric haloscope, such as MADMAX~\cite{Caldwell:2016dcw, Millar:2016cjp, MADMAX:2019pub}, as a high frequency GW detector. Dielectric haloscopes consist of a stack of thin disk with large dielectric constant placed in a static external magnetic field. A coherent non-relativistic dark matter axion field sources electromagnetic (EM) waves at the surfaces of these disks. These waves propagate orthogonal to the disk surfaces until they reach a receiver placed at one end of the stack. If the disk separation is tuned to the axion mass, the electromagnetic waves sourced at the different disks interfere constructively, boosting the signal by the number of disks.

Remarkably, while this technology promises world-leading bounds for axions in the frequency range 10--\SI{100}{GHz} in the next years, it turns out to be arguably better-suited to search for signals from relativistic sources, such as gravitational waves, in a similar frequency range. 
This can be traced back to two different resonant phenomena. First, similar to the axion case, the detector geometry including in particular the disk separation and thickness, can be tuned to obtain constructive interference between EM waves sourced at the surface of the different disks. We will refer to this setup as the resonant operation regime of the dielectric haloscope. The specific requirements on disk placement, thickness and smoothness differ from the axion case and will be discussed in detail. Second,  since both the GW and the photon are massless, the GW-to-photon conversion in vacuum (i.e.\ between the disks) occurs on resonance~\cite{Raffelt:1987im} at all frequencies -- photons generated at different spacetime points interfere constructively. Compared to the axion case, this leads to an additional $\omega \ell \gg 1$ enhancement of the signal, where $\omega$ is the frequency of the GW and $\ell$ the detector length. This leads to competitive sensitivities operating in a broadband configuration without dielectric disks, relying exclusively on the GW-to-photon conversion in vacuum. As a result, we will consider three different operation modes: the resonant and broadband modes mentioned above, as well as a hybrid resonant/broadband mode.

In the following, we first calculate in \cref{sec:EMsolutions} the oscillating electromagnetic (EM) field induced by a gravitational wave in vacuum and in a medium with a dielectric constant $\epsilon> 1$. In \cref{sec:TransferMatrix} we then establish the transfer matrix formalism we use to calculate the magnitude of the EM fields anywhere inside a dielectric haloscope apparatus. The formalism allows us to take into account all reflection and interference effects. We use these results in \cref{sec:ResonantMode} to estimate the magnitude of the GW-induced EM flux reaching a receiver in a fully-resonant setup. We then demonstrate in \cref{sec:Hybrid} how a particular placement of the dielectric disks allows for performing a simultaneous resonant and broadband experiment covering different GW frequency bands, and we discuss the possibility of disk-less (broadband-only) operation. In \cref{sec:sensitivity} we estimate the expected sensitivity of MADMAX in the various different operational modes, before concluding in \cref{sec:conclusion}. Further details are provided in the appendices. \Cref{app:conventions} lists our normalization conventions, \cref{app:TT-HFlimit} comments on the use of the transverse traceless frame in the high-frequency limit, \cref{app:transfermatrix} provides further details on the derivation of key quantities entering the transfer matrix formalism, \cref{app:2angles} discusses the full angular dependence of our solutions and finally \cref{app:PBHdistance} discusses the sensitivity of our proposed setup to mergers of light primordial black holes.

\section{GW induced electromagnetic fields}
\label{sec:EMsolutions}

\begin{figure}
    \center
    \includegraphics[width=\columnwidth]{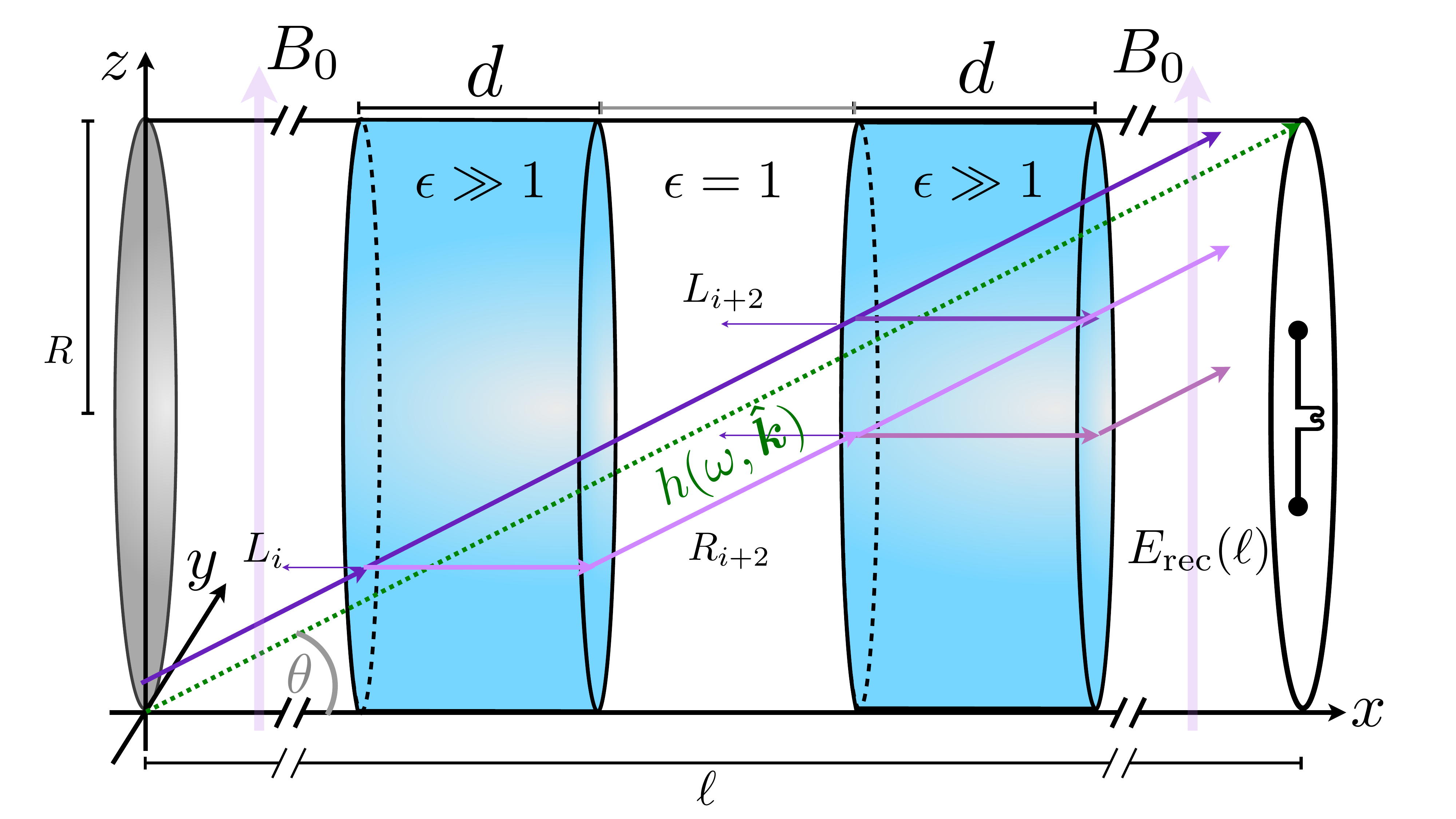}
    \caption{Schematic setup of a dielectric gravitational wave detector of length $\ell$ with only two disks (light blue). The thickness of the disks is inflated here to show more clearly the trajectories inside. In reality, $d \ll \ell, \, R$. The incoming GW is depicted as dotted green line, the sourced electromagnetic waves are shown in solid purple. At each disk surface, boundary conditions enforce the presence of both left-moving ($L_i$) and right-moving ($R_i$) EM plane waves. The full field at the receiver $E_{\rm rec}(\ell)$ is the sum of all the right-moving waves plus the particular solution of Maxwell's equations in the presence of the GW source. We use the transfer matrix formalism described in \cref{sec:TransferMatrix} to compute the relevant quantities. The receiver is depicted as a dipole antenna at the right end of the setup.}
    \label{fig:sketch_pws}
\end{figure}

Our proposal closely follows the concept of dielectric axion haloscopes such as MADMAX~\cite{Caldwell:2016dcw, Millar:2016cjp, MADMAX:2019pub}, which consist of a stack of dielectric disks placed inside a static magnetic field $\vec{B}^e = B_0 \vec{\hat e}_z$ parallel to the disk surfaces. A receiver is placed at one end of the stack in a field-free region. A two-disk cartoon of the setup is shown in \cref{fig:sketch_pws}, with the thickness of the disks inflated for clarity. In close analogy to the axion, an incoming GW sources an effective current~\cite{Gertsenshtein:1962, Boccaletti:1970pxw, Berlin:2021txa, Domcke:2022rgu}, which can be expressed in terms of an effective magnetization and polarization~\cite{Domcke:2022rgu},
\begin{align}
  \vec{j}_\text{eff} = \vec{\nabla} \times \vec{M}_\text{eff} + \vec{\dot P}_\text{eff} \,.
  \label{eq:jeff}
\end{align}
For gravitational waves, $\vec{P}_\text{eff} = 0$ and $\vec{M}_\text{eff} = - h_{ij}^{TT} B_0^j$. Here we work in the transverse traceless (TT) frame for the gravitational wave, where the metric perturbation can be written as $h_{ij}^{TT} = (h^+ \hat e^+_{ij} + h^\times e^\times_{ij}) \, e^{- i \omega (t-\boldsymbol{\hat {k}} \cdot \boldsymbol{x})}$ with the unit tensors $\hat e^+_{ij}$ and $\hat e^\times_{ij}$ satisfying $\hat e^\lambda_{ij} \hat e_{\lambda'}^{ji} = 2 \delta_{\lambda \lambda'}$. Choosing the TT frame is convenient since we will be considering frequencies far above the relevant mechanical resonance frequencies of the experimental setup, such that all constituents of the setup can be considered as free-falling in their response to GWs, see \cref{app:TT-HFlimit} for details. Maxwell's equations now read
\begin{align}
  \vec\nabla \cdot \vec{E} & = 0 \,, \quad \vec\nabla \times \vec{B} - \epsilon \vec{\dot E} = \vec{j}_\text{eff} \,,
  \label{eq:MW_1}\\
  \vec\nabla \cdot \vec{B} & = 0 \,, \quad \vec\nabla \times \vec{E} + \vec{\dot B} = 0 \,, 
  \label{eq:MW}
\end{align}
where $\epsilon$ is the dielectric constant and we have set the magnetic permeability ($\mu$) to unity. Note that these equations apply to isotropic materials. If one were to consider anisotropic materials, new terms in the inhomogeneous Maxwell's equations would appear at $\mathcal{O}(h)$ due to the effect of the GW on the dielectric tensor. While the LaAlO$_3$ perovskite the MADMAX collaboration plans to use for their disks~\cite{MADMAX:2019pub} is mildly anisotropic~\cite{makeev2002anisotropy, delugas2005dielectric}, we leave the detailed investigation of the effect of such anisotropies to future work.

Assuming an isotropic material, we obtain a particular solution to these equations for $\epsilon \neq 1$ and $\vec{\hat{k}}$ in the $x$--$z$ plane,\footnote{The more general solution for a GW with arbitrary incoming direction for which two angles, $\theta$ about the $y$-axis and $\phi$ about the $z$-axis, are required is given in \cref{app:2angles}.}
\begin{align}
    \vec{E}^p_m = \frac{c_\theta B_0}{\epsilon - 1} \, (h_\times \vec{\hat p} + h_+ \vec{\hat s}) \,
                  e^{- i \omega (t - \vec{\hat k} \cdot \vec{x})} \,,
    \label{eq:3D_med_p}
\end{align}
Here $\theta$ is the angle between the incoming GW and the symmetry axis of the detector,  $c_\theta = \cos \theta = \vec{\hat k} \cdot \vec{\hat e}_x$.\footnote{In general $\theta$ can take on any value, but since the MADMAX apparatus will be instrumented at one end only, we restrict ourselves to considering solutions with $|\theta|<\pi/2$.}
The unit vectors $\vec{\hat s} = \vec{\hat e}_y$ and $\vec{\hat p} = \vec{\hat k} \times \vec{\hat s}$ denote the direction of the $s$ (``senkrecht'' meaning ``orthogonal'' to the plane of incidence) and $p$ (``parallel'' to that plane) polarized components of the electric field. The corresponding magnetic field is obtained immediately from \cref{eq:MW}. In vacuum, $\epsilon = 1$, we find the particular solution
\begin{align}
    \vec{E}^p_v = - \frac{B_0}{2} \left[ i \omega x  (h_\times \vec{\hat p} + h_+ \vec{\hat s}) +  h_\times s_\theta \vec{\hat k} \right]                                                  
    e^{- i \omega (t - \vec{\hat k} \cdot \vec{x})} \,.
    \label{eq:3D_vac_p}
\end{align}
One can immediately verify that \cref{eq:3D_med_p,eq:3D_vac_p} solve \cref{eq:MW_1,eq:MW}.
To obtain the solutions in \cref{eq:3D_med_p,eq:3D_vac_p}, we used an ansatz proportional to an exponential function reflecting the time and space dependence of the source term $\vec{j}_\text{eff}$, allowing in the vacuum case for a prefactor proportional to $x$ to account for the continuous (resonant) GW to photon conversion, which happens all along the GW trajectory, and is proportional to the background magnetic field component perpendicular to the GW propagation.
In practice, resonant conversion will be limited by an imperfect vacuum, or ultimately the Euler--Heisenberg contribution to the photon mass and the backreaction of the electromagnetic wave on the GW wave equation~\cite{Raffelt:1987im}. All of these effects can be shown to be irrelevant for a meter-scale experiment.
It is important to keep in mind that \cref{eq:3D_med_p,eq:3D_vac_p} are not unique -- one can always add plane waves (i.e.\ solutions to the free Maxwell equations) without loss of generality, see \cref{app:transfermatrix} for details.

\section{Transfer Matrix Formalism}
\label{sec:TransferMatrix}

In order to match the boundary conditions at the surface of the dielectric disk, $\vec{B}_m = \vec{B}_v$,  $\vec{E}^\parallel_m = \vec{E}^\parallel_v$ and $\epsilon \vec{E}^\perp_m = \vec{E}^\perp_v$,\footnote{Here, the superscript $\parallel$ denotes the component of the electric field parallel to the disk surfaces, and $\perp$ stands for the component perpendicular to them. The subscript $m$ and $v$ are for quantities in medium ($\epsilon > 1$) and vacuum, respectively.} we need to add the appropriate plane-wave solutions sourced at the disk surfaces to the particular solutions from \cref{eq:3D_med_p,eq:3D_vac_p}. For a system with multiple boundaries, this is most conveniently implemented using the transfer matrix formalism. In the following, our notation follows closely the corresponding derivation for the axion case in Ref.~\cite{Millar:2016cjp}, which we apply to the $s$ and $p$-polarized EM waves separately.

We divide the setup into $2N_d + 1$ regions of constant $\epsilon$, with the first region corresponding to the vacuum to the left of the leftmost disk, the second region corresponding to the interior of the first disk, and so on. In each region, the EM field is described by a 2-component vector $E_n = (R_n, L_n)$, with the upper component describing the amplitude of right-moving EM waves and the lower component that of left-moving EM waves at the left edge of the $n$-th-region. Focusing first on propagation, reflection, and transmission of EM waves in the absence of GWs, the $E_n$ are acted upon by $2 \times 2$ matrices that describe propagation ($P_{v,m}$) through a vacuum or medium region and matching across a vacuum-to-medium boundary ($G_{mv}$) or medium-to-vacuum boundary ($G_{vm}$). (In what may be a confusing convention, the second subscript stands for the region on whose $E_n$ the matrix is acting, and the first one for the region it maps onto.) Explicit expressions for the propagation matrices are obtained from the free Maxwell equations in vacuum and medium, and are given in \cref{eq:prop}. The transfer matrices $G_{mv,vm}$ are found by solving the boundary conditions at the disk surfaces, with the results listed in \cref{eq:Gmv,eq:Gvm}.  In the transfer matrix formalism, an EM wave propagating through a single detector segment, containing a vacuum region of length $D$ followed by a disk of width $d$ evolves according to
\begin{align}
    E_{n + 2} = \left( G_{vm} \cdot P_m \cdot G_{mv} \cdot P_v \right) \cdot E_n \,.
\end{align}
The quantity in brackets, $T_1 \equiv G_{vm} P_m G_{mv} P_v$, can be interpreted as an effective single-segment transfer matrix that describes transmission and reflection on that segment. For a setup with $N_d$ segments, the plane wave amplitudes at the right-hand edge of the device are related to those on the left-hand edge by the $N_d$-th power of the single-segment transfer matrix, $E^{N_d} = T_1^{N_d} E_0$. 

The transmission ($\mathsf{t}$) and reflection ($\mathsf{r}$) coefficients for right-moving and left-moving EM plane waves can therefore be defined as
\begin{align}
    &\mathsf{t}_R \equiv \frac{1}{\big(T_1^{N_d}\big)_{22}} \,,
     \qquad
     \mathsf{t}_L \equiv \frac{\text{det}\big(T_1^{N_d}\big)}{\big(T_1^{N_d}\big)_{22}} \,,
    \label{eq:transmissionCoeffs} \\[0.2cm]
    &\mathsf{r}_R \equiv \frac{\big(T_1^{N_d}\big)_{12}}{\big(T_1^{N_d}\big)_{22}} \,,
     \qquad
     \mathsf{r}_L \equiv -\frac{\big(T_1^{N_d}\big)_{21}}{\big(T_1^{N_d}\big)_{22}} \,.
    \label{eq:reflectionCoeffs}
\end{align}

The EM waves resulting from GW conversion in the various detector components appear as additional source terms in the transfer matrix formalism. Solving for the boundary conditions at the disk surfaces including the particular solutions~\eqref{eq:3D_med_p} and \eqref{eq:3D_vac_p} determines the amplitudes of the plane waves on the right-hand side of the surface `sourced' by the GW. This is encoded in source vectors $S_{mv}^{(n)}$ and $S_{vm}^{(n)}$ for vacuum-to-medium and medium-to-vacuum interfaces, respectively. For a GW incoming along the symmetry axis of the detector, $\theta = 0$, the resulting source terms at the $n$th disk are
\begin{align}
    &S^{p,s}_{mv} =\!\frac{B_0 h_{\times,+}}{4 \sqrt{\epsilon}} e^{i x_n \omega} \,
                   \frac{\sqrt{\epsilon} + 1}{\sqrt{\epsilon} - 1}\!
                   \begin{pmatrix}
                       (1 \!+\! i x_n (\sqrt{\epsilon}-1) \omega) \frac{\sqrt{\epsilon}+1}{\sqrt{\epsilon}-1} \\
                       (1 \!-\! i x_n (\sqrt{\epsilon}+1) \omega) \frac{\sqrt{\epsilon}-1}{\sqrt{\epsilon}+1}
                   \end{pmatrix} \,,
    \label{eq:sourceVacMed}
\intertext{and}
&S^{p,s}_{vm} = \frac{B_0 h_{\times,+}}{4} e^{i (x_n + d) \omega}
                   \begin{pmatrix}
                       1 + 2 i (x_n + d) \omega + \frac{4}{\epsilon - 1}) \\
                       -1
                   \end{pmatrix} \,,
    \label{eq:sourceMedVac}
\end{align}
with $x_n = n D + (n - 1) d$ denoting the position of the left edge of the $n$th disk. The generalization to generic angles $\theta$ is straightforward, and the corresponding expressions are given in Appendix~\ref{app:transfermatrix}.

It is worth highlighting the major differences between these source terms and those obtained for axion dark matter in Ref.~\cite{Millar:2016cjp}. The first, most notable difference is the appearance of the phase factor $e^{i x \w}$, absent for non-relativistic axions,\footnote{Although note that the small axion velocity results in a similar factor, see Ref.~\cite{Millar:2017eoc}.} where $x$ is the position of the interface. This reflects the fact that the GW is propagating through the detector at the speed of light.  The second notable difference is that besides the order-unity terms in the source vectors, there are additional position-dependent terms scaling as $x\,\w$. Their appearance reflects the resonant conversion between the massless GW and photon in vacuum all along the GW trajectory, first seen in the particular solution of Maxwell's equations in \cref{eq:3D_vac_p}.

The effective source term including both surfaces of a single disk can be expressed as
\begin{align}
    M^{(n)} = S_{vm}^{(n)} + G_{vm} \cdot P_m \cdot S_{mv}^{(n)}\,.
\end{align}
The appearance of the in-medium propagation matrix $P_m$ multiplying the vacuum-to-medium source vector, combined with the relative phase $e^{id\w}$ between the two source vectors \cref{eq:sourceVacMed,eq:sourceMedVac} means that $M^{(n)}$ depends non-trivially on the thickness of the disks. This implies a requirement on the disk thickness to ensure constructive interference between the right-moving EM waves sourced at the two surfaces: for a generic incoming GW angle in the $x$--$z$ plane, we find that $M^{(n)}$ is maximized for a disk thickness that satisfies
\begin{align}
    d_{\rm max} \sqrt{\epsilon} \cos\theta_\epsilon = \pi/(2 \omega) \,.
    \label{eq:dmax}
\end{align}
The quantity $\theta_\epsilon$, defined by $\epsilon \sin\theta_\epsilon = \sin\theta$ accounts for the refraction of the EM plane wave in the medium.  If \cref{eq:dmax} is satisfied, we expect from \cref{eq:3D_vac_p} that the source terms at the $n$th disk contribute EM waves with an amplitude of order $\omega x_n h B_0$.

Combining the transfer matrix propagating EM plane waves with the effective source term from the GW enables us to express the amplitude of the EM waves at the receiver, placed a distance $\ell$ away from the left edge of the device, as
\begin{align}
    E_\text{rec}(\ell) = P^\ell_v \Big( T_1^{N_d} E_0 + \sum_{n = 1}^{N_d} T_1^{N_d-n} M^{(n)} \Big) + E_p(\ell)\,.
    \label{eq:ErecE0}
\end{align}
Here, $E_0 = (0, L_0)$ denotes the amplitude of left- and right-moving plane waves at the left edge of the setup. The second term in parentheses encodes the contribution sourced at the disks and then propagated through the subsequent disks to the receiver. The additional factor $P^\ell_v$ accounts for the propagation in vacuum after the last disk, through a distance $x_{\rm eff} = \ell\,c_\theta - N_d(D\,c_\theta+d\,c_{\theta_\epsilon}) \geq 0$ which need not be equal to the effective length $D\,c_\theta$ of the other vacuum sections.  Finally, the last term $E_p(\ell)$ adds the amplitude of the right-moving particular solution, \cref{eq:3D_vac_p}, at the position of the receiver. It corresponds to the signal that would be received in the absence of the dielectric disks.

\begin{figure}
    \centering
    \includegraphics[width=0.9\linewidth]{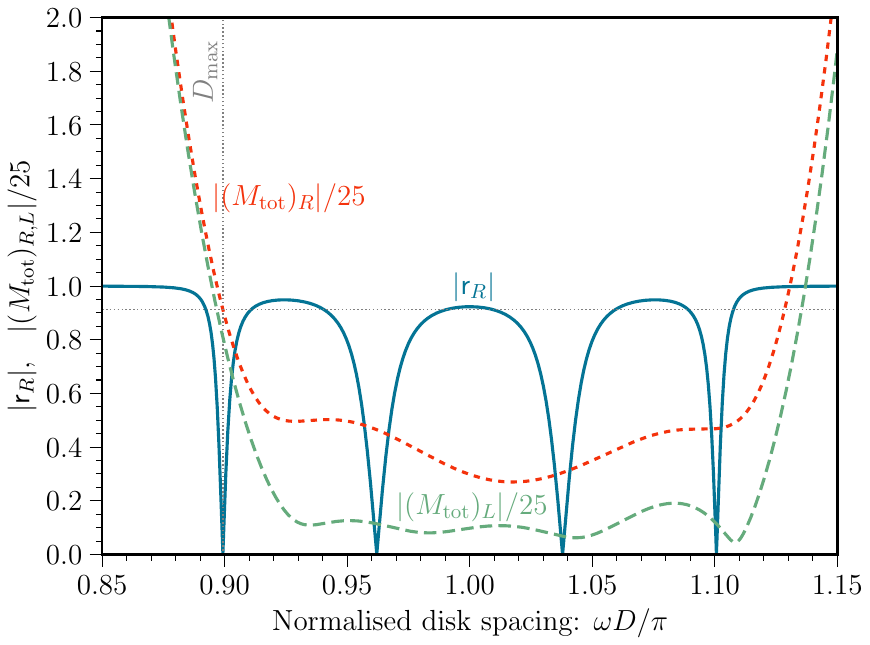}
    \caption{The reflection coefficient $|\mathsf{r}_R|$ (blue solid) and effective source terms $|(M_{\rm tot})_R|/25$ (red dotted) and $|(M_{\rm tot})_L|/25$ (green dashed) as a function of $D$ in units of $\w/\pi$ for $N_d = 5$ and $\epsilon = 25$. We see that the analytical result \cref{eq:lmax}, which corresponds to the dotted vertical line, also corresponds to the largest value of $(M_{\rm tot})_R$ that coincides with a zero of $|\mathsf{r}_R|$ (horizontal grid line). From \cref{eq:ErecDet}, we see that this value of $D$ maximises the received electric field.}
    \label{fig:lmax}
\end{figure}

We now impose boundary conditions at the two edges of the detector, in particular
a vanishing amplitude for the incoming left-moving EM wave at the right edge of the setup, $E_\text{rec} = (E_\text{rec}^\text{R},0)$, and a vanishing amplitude for the incoming right-moving EM wave at the left edge of the setup $E(x=0) = (0,L_0)$. This neglects, for simplicity, the possible presence of  a mirror at the left edge of the apparatus.
We can now solve for $L_0$ and the amplitude of the total right-moving wave at the receiver, $E_\text{rec}^\text{R}$. Denoting $M_\text{tot} = \sum_{n = 1}^{N_d} T_1^{N_d-n} M^{(n)}$, the resulting solution for the received electric field can be written compactly as
\begin{align}
    \label{eq:ErecDet}
    E_\text{rec}^\text{R}(\ell) = E_p(\ell)+ e^{i \w x_{\rm eff}}\big[ \left(M_\text{tot}\right)_R - \left( M_\text{tot}\right)_L \mathsf{r}_R \big] \ ,
\end{align}
where $\mathsf{r}_R$ is the reflection coefficient for right-moving EM waves given in \cref{eq:reflectionCoeffs}. The subscripts $\{R,L\}$ indicate the right- and left-moving components of the $N_d$-disk source vectors, respectively. \Cref{eq:ErecDet} enables us to further understand the requirements for resonant enhancement of GW signals in a dielectric haloscope. Examining the second term of \cref{eq:ErecDet}, it is clear that for certain combinations of signs for the three quantities in square brackets, partial or full cancellation occurs (see also \cref{app:transfermatrix} for a more detailed discussion). We can impose the condition $\mathsf{r}_R = 0$ to avoid this cancellation. Physically, minimizing the reflection coefficient for right-moving waves in this way is equivalently to maximizing the transmission coefficient, thereby maximizing the EM signal at the receiver. Using \cref{eq:reflectionCoeffs}, the condition $\mathsf{r}_R = 0$ can be translated into a condition on the separation of the disks, $D$. For a GW propagating along the detector axis, this condition is found to be
\begin{align}
    \label{eq:lmax}
    D_{\rm max}\, \w = k\pi-\text{arcsin}\left( \frac{2\sqrt{\epsilon} \cos (\pi/N_d)}{1+\epsilon}\right) \ .
\end{align}
Note that this solution is not unique. However, among all solutions of the equation $\mathsf{r}_R(D) = 0$, \cref{eq:lmax} is the one with the largest $(M_\text{tot})_R$, thus maximizing the signal at the receiver. This analytical result for the optimal disk separation is validated numerically in \cref{fig:lmax} for a 5-disk system with $k=1$ and $\epsilon = 25$ (see also Fig.~\ref{fig:resonances} in \cref{app:transfermatrix}). We find that our analytic estimate is an excellent approximation to the numerically-obtained $D_{\rm max}$, disagreeing only at $\mathcal{O}(10^{-5})$ for $N_d = 10$, and that the difference scales as $1/N_d^4$. As a result, for large $N_d\gtrsim 10$, the analytic $D_{\rm max}$ is sufficient to estimate the effect of imperfections in disk placement, as we do in the next section.

For a GW with a general incident angle \cref{eq:lmax} will be modified. This implies that for a fixed disk width, changing the disk separation amounts to scanning through some combination of GW frequencies and incident angles. Even though this means that for a given GW frequency, a dielectric haloscope will have an extremely narrow field of view, this is not a severe limitation in practice, given that realistic signals either scan over a range of frequencies (e.g.\ primordial black hole mergers) or have a broad frequency spectrum. The details of the frequency and angle-dependent response functions can be determined numerically using the formalism described here.

\section{Detecting Gravitational Waves in Resonant Mode}
\label{sec:ResonantMode}

Imposing the optimal disk thickness from \cref{eq:dmax} and disk spacing from \cref{eq:lmax}, we can estimate the maximal electric field at the receiver as
\begin{align}
    E_\text{rec}^\text{R} \sim \sum_n \omega x_n h B_0 \sim N_d \omega \ell h B_0 \, ,
    \label{eq:ErecR}
\end{align}
which holds for moderately large $N_d$. This approximate result arises due to the domination of the disk-sourced EM plane waves over the particular solution in \cref{eq:ErecR}. It allows us to rapidly estimate the scaling of the electromagnetic flux density into the receiver, given by the time-averaged Poynting vector
\begin{align}
    \vec{S} = \frac{1}{2}\text{Re}[\vec{E}_{\rm rec}^R \times (\vec{B}_{\rm rec}^R)^*]
              \sim  h^2(N_d\,\w\,\ell\,B_0)^2 \,.
    \label{eq:Poynting}
\end{align}
The factor $1/2$ arises in the time-averaging due to the use of phasor notation.

It is instructive to compare this expression to the axion case, $\vec{S} \sim (g_{a \gamma} a B_0 N_d)^2$ with $a$ the axion amplitude and the boost factor $N_d^2$ enhancing the signal for a large number of disks. Comparing the couplings to the EM fields in the Lagrangian, $g_{a\gamma} a F \tilde F$ for axions and $hF^2$ for GWs, one might have expected to obtain the parametric result for GWs by simply replacing $g_{a \gamma} a \mapsto h$. Instead, we note the
additional $(\omega \ell)^2$ enhancement due to resonant GW-to-photon conversion in vacuum.\footnote{See \cref{app:TT-HFlimit} for a discussion on how this frame-independent result can be understood in other reference frames.} We stress that the latter is present also in the absence of any disks (as has been noted e.g.\ in the context of GW searches with helioscopes and light-shining-through-wall experiments~\cite{Ejlli:2019bqj, Ringwald:2020ist}). In this case the Poynting vector scales as $\vec{S}_{\rm vac} \sim h^2 (\w\,\ell\,B_0)^2$, indicating that the presence of disks leads to a boost factor $|\vec{S}|/|\vec{S}_{\rm vac}| \sim N_d^2$, in this sense similar to the axion case.  Note that this scaling of the boost factor only holds for a moderate number of disks, and for perfect disk placement and disk thickness. Below we discuss in greater detail how large a disk number is beneficial, including how it depends on the dielectric constant of the disks. We also discuss the impact of imperfections, both in disk placement and thickness.

\subsection{$N_d$-dependence of the sensitivity for variable apparatus length}

\begin{figure}
    \centering
    \begin{tabular}{c}
        \includegraphics[width=\columnwidth]{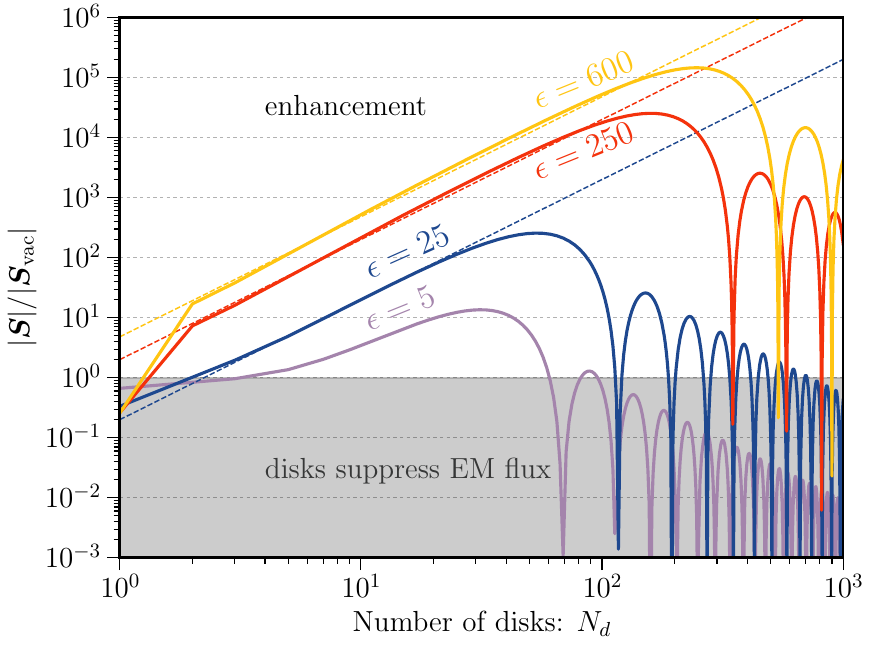} \\
        \includegraphics[width=\columnwidth]{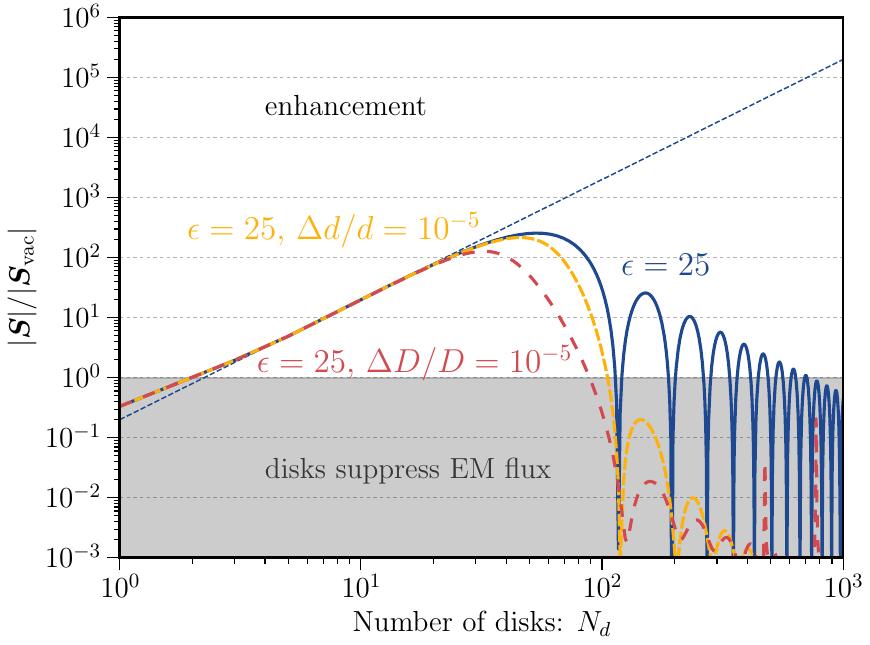}
    \end{tabular}
    \caption{Performance of a dielectric haloscope operated in resonant mode with \emph{variable} detector length $\ell = (N_d+1) D_{\rm max} + N_d \, d_{\rm max}$.
    \textbf{\textit{Upper panel:}} Poynting vector ratio $|\vec{S}|/|\vec{S}_{\rm vac}|$ as a function of the number of disks for perfect disk placement ($D_\text{max}$) and thickness ($d_\text{max}$). Different colored curves correspond to different values of the dielectric constant $\epsilon$. The curves demonstrate the $\vec{S} \propto N_d^2$ growth until a plateau is reached for $N_d \gtrsim 10\sqrt{\epsilon}$.
    \textbf{\textit{Lower panel:}} Fixing $\epsilon = 25$, we study the impact of imperfect disk placement (red curve) and imperfect disk thickness (yellow curve) compared to case of the perfect $D_\text{max}$ and $d_\text{max}$ (blue). Also shown is a polynomial fit $\vec{S}\propto N_d^2$ in dashed blue.
    }
    \label{fig:PoyntingVariableLength}
\end{figure}

At large $N_d$, the growth in the received electric field saturates since the EM waves sourced at two disks that are separated by a larger number of disks no longer interfere constructively. Technically, this can be seen by evaluating the effective source term $(M_\text{tot})_R$ for optimized disk width $d_\text{max}$ and disk separation $D_\text{max}$ and noticing that the initial quadratic growth with $N_d$ (see \cref{app:transfermatrix}) is replaced by an oscillating function of $N_d$.
Empirically, we find that this saturation occurs for $N_d \gtrsim 10 \sqrt{\epsilon}$. For even larger values of $N_d$, the EM plane waves sourced at the disk surfaces and the particular solution $E_p(\ell)$ interfere destructively, and the value of the field no longer grows with the detector length $\ell$, either. This destructive interference can happen because for fixed $d$ and $D$, larger $N_d$ also means larger $\ell$, enhancing $E_p(\ell)$. This can be seen in \cref{fig:PoyntingVariableLength}, where we plot the absolute value of the Poynting vector, $|\vec{S}|$, relative to its vacuum value, $|\vec{S}_{\rm vac}|$, as a function of the number of disks (and thereby also as a function of the apparatus length $\ell = (N_d+1) D_{\rm max} + N_d \, d_{\rm max}$. We see that at moderate $N_d$, the colored curves scale as $|\vec{S}|/|\vec{S}_{\rm vac}| \propto N_d^2$, which, given that $|\vec{S}_{\rm vac}| \propto \ell^2$, demonstrates the $N_d^2 \ell^2$ growth of the disk-enhanced solution expected from \cref{eq:ErecR}. For very small number of disks, or very small $\epsilon$, the scaling differs due to the contribution from the particular solution $E_p(\ell)$. At large $N_d$, the curves develop the aforementioned plateau and then begin to oscillate with $N_d$. Comparing different values of the dielectric constant $\epsilon$ (different coloured curves in the upper panel of \cref{fig:PoyntingVariableLength}), we observe that the maximum number of disks before saturation is reached is dependent on $\epsilon$. As stated above, we find empirically that $N_d \gtrsim 10\sqrt{\epsilon}$ is a good approximation to the saturation point. Evidently, a material with as large a dielectric constant as possible is preferred, modulo difficulties with manufacturing.

\subsection{Impact of disk imperfections}

To demonstrate the effect of detector imperfections, in the lower panel of \cref{fig:PoyntingVariableLength} we show in yellow the Poynting vector for a setup with perfect disk placement, but imperfect disk thickness $d = d_{\rm max} \left(1+10^{-5}\right)$, and in red the solution with perfect disk thickness, but imperfect placement by the same relative amount. Evidently, there is a greater tolerance to imperfections in the disk thickness than in the disk placement, as the latter leads to a more significant reduction in both the peak value of $|\vec{S}|/|\vec{S}_{\rm vac}|$ and in the maximal useful number of disks.\footnote{Note that in the previous section, we discussed the deviation of the numerical $D_{\rm max}$ from the analytic expression given in \cref{eq:lmax}. We found that for $N_d \sim 10$, the analytic estimate was accurate at the $10^{-5}$ level, and its accuracy improved at larger $N_d$. Since \cref{fig:PoyntingVariableLength} shows that the effect of an imperfection in the disk placement of a similar order is only important upon the placement of $N_d \gtrsim 30$ disks, our estimates here are well within the domain of validity of \cref{eq:lmax}.}

Requirements on accurate and tunable disk separation $D$ are an integral part of the axion program in dielectric haloscope experiments. A variable effective disk width $d \sqrt{\epsilon} \cos \theta_\epsilon$, on the other hand, would be a new requirement of a tunable resonant GW search. As manufacturing and installing a new stack of disks at each step in a scan over frequencies might be prohibitively difficult, it may in practice be easier to vary not $d$, but the dielectric constant, e.g.\ by filling the disks with some gas or liquid or by the use of metamaterials~\cite{ALPHA:2022rxj}. Despite these difficulties, we will in the remainder of this paper show results for fully or partially resonant operation assuming $d = d_\text{max}(f)$ to illustrate the maximal possible reach of this approach. On the other hand, since in the GW case the propagation direction of the EM waves is set by the GW wave vector, whereas in the axion case it is given by the orientation of the disk surface, the disk smoothness requirements are less stringent here than in the axion case.

\subsection{Sensitivity for fixed apparatus length}

In the discussion so far, adding more disks to the apparatus always meant extending its length. However, in a realistic setup, the length is fixed, and one must place disks in the available space. One option is to place all disks at the optimum separation $D_{\rm max}$ as before (with $k=1$ in \cref{eq:lmax}) such that the disks only occupy a portion of the apparatus length (preferably on the side furthest from the receiver). In this case the device can be operated in a hybrid resonant/broadband mode, which will be discussed further in the next section. Here, we focus instead on another option, namely choosing $k = k_{\rm fill}$ in \cref{eq:lmax}, where $k_{\rm fill}$ is the integer that spaces the disks so as to fill as much of the apparatus length as possible with evenly separated disks. For fixed $\ell$ and $N_d$, we can use \cref{eq:lmax} to determine $k_{\rm fill}$, finding
\begin{align}
    \label{eq:kfill}
    k_{\rm fill}\! =\! \text{floor}\! \left[
                                    \text{arcsin}\!\left( \frac{2\sqrt{\epsilon} \cos (\pi/N_d)}{1+\epsilon} \right)
                                  \!+\! \frac{2 \ell \w  - \frac{N_d \pi}{\sqrt{\epsilon}}}{2\pi(1+N_d)}
                                \right] \,.
\end{align}
This expression is dominated by the second term balancing the apparatus length with the number of disks. For the design parameters set out by the MADMAX collaboration in Ref.~\cite{MADMAX:2019pub}, $\ell = \SI{2}{m}$ and $\epsilon = 25$, targeting the frequency range $\w \in 2 \pi [10^{10}, 10^{11}]\,\si{Hz}$, we find that there is a maximum number of disks $N_d \lesssim 136 \, [\w / (2\pi \times \SI{e10}{Hz})]$ that can be placed in the apparatus when targeting a resonance frequency $\w$.

\begin{figure}
    \centering
    \includegraphics[width=\columnwidth]{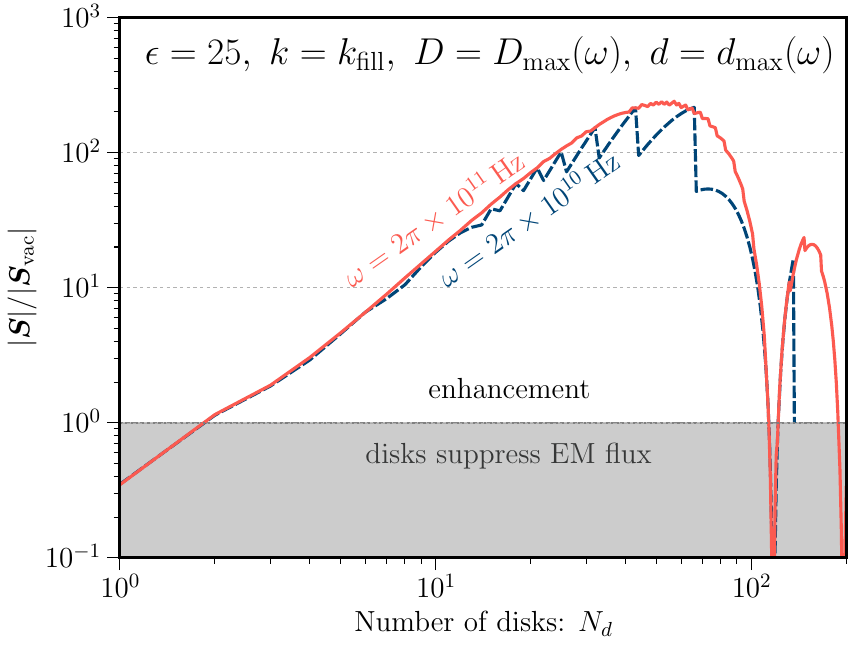}
    \caption{Performance of a dielectric haloscope operated in resonant mode for \emph{fixed} detector length $\ell = \SI{2}{m}$, choosing $k = k_{\rm fill}$ in the solution for $D_{\rm max}$, so that the full length of the apparatus is filled with disks. As in \cref{fig:PoyntingVariableLength}. we plot Poynting vector ratios $|\vec{S}|/|\vec{S}_{\rm vac}|$ as a function of $N_d$. Different colored curves correspond to different values of $\w$. As discussed in the text, for $\w = 2 \pi \times \SI{e10}{Hz}$ up to $N_d = 136$ disks can fit into the detector, explaining the abrupt end of the blue curve at this value. The jaggedness of both curves comes from the requirement that $k_{\rm fill}$ must be an integer number. We observe that the maximum Poynting vector ratio is $|\vec{S}|/|\vec{S}_{\rm vac}| \sim 200$, obtained for $40 \lesssim N_d \lesssim 70$.
    }
    \label{fig:maxNdkfiill}
\end{figure}

While this shows that the maximum number of disks that can be placed in the apparatus can be large, what is more important is whether placing additional disks actually leads to an increased signal flux at the receiver. From the results of the previous section, we can already anticipate that the optimal \emph{useful} number of disks will be significantly smaller than the number that could in principle fit in the apparatus. To evaluate the optimal useful number of disks, we compute the ratio $|\vec{S}|/|\vec{S}_{\rm vac}|$, setting $k=k_{\rm fill}$ in $D_{\rm max}$ and fix $\ell = \SI{2}{m}$, $\epsilon = 25$. The result is shown in \cref{fig:maxNdkfiill} for two different values of $\w$. We see that as in the previous discussion where we allowed the detector length $\ell$ to be a function of $N_d$, fixing $\ell$ and maximising $k$ by setting it equal to $k_{\rm fill}$ yields almost the same result for the maximum number of useful disks, $N_d \sim 60$ for $\epsilon = 25$. Of note is that the two coloured curves in \cref{fig:maxNdkfiill} follow the same shape, implying that the ratio of Poynting vectors is independent of the signal frequency when all quantities $D$, $d$, $k$ are optimised. The $\w = 2\pi \times \SI{e10}{Hz}$ curve ends at $N_d = 136$ as this is the maximum number of disks that can fit into the \SI{2}{m}-long detector at this frequency while keeping $d$ and $D$ optimal. Finally, recalling the previous discussion, we note that using a material with a larger $\epsilon$ leads to a larger number of useful disks, scaling as $N_d \propto \sqrt{\epsilon}$.

We conclude that in fully-resonant mode, optimising the disk spacing leads to an almost frequency-independent maximum signal flux enhancement for the future MADMAX detector of $|\vec{S}|/|\vec{S}_{\rm vac}| \sim 200$, which in turn translates to a maximum enhancement of the strain sensitivity relative to a setup without disks of $h/h_{\rm vac} \sim 14$. This enhancement is not insignificant, and can be improved by using disks with larger dielectric constants. However, it comes at the price of rendering the device fully resonant, with fairly stringent tolerances on the disks. As a result, in the next section we consider two alternative operation modes for MADMAX: one in which there are disks but also a large vacuum region, and one in which the disks are entirely removed.

\section{Hybrid Resonant/Broadband or Pure Broadband Operation}
\label{sec:Hybrid}

As discussed in the previous section, fully resonant operation of MADMAX can lead to an enhancement in the strain sensitivity to GWs that can be greater than an order of magnitude compared to a setup without dielectric disks. However, given the nature of expected sources of high-frequency GWs such as sub-solar mass BH mergers, it can be beneficial to have broadband sensitivity to access a range of frequencies simultaneously. We consider two options for achieving this here: a hybrid resonant/broadband setup; and a fully broadband setup without any dielectric disks. In the former, the stack of dielectric disks (whose thickness and spacing is optimised for a particular frequency) is made as compact as possible by choosing $k=1$ in \cref{eq:lmax} and then placed at the end of the device furthest from the receiver.

Consider first the hybrid setup. It has the advantage that GWs whose wave vector points in a direction such that waves pass through the disks and the receiver are resonantly enhanced by the disks, provided their frequency matches the setup's resonance frequency. The signals from GWs at other frequencies experience less enhancement, or even suppression from the disks. However, for a certain range of incoming GW directions, the GW wave vector does not pass through the disks and the receiver simultaneously. GWs hitting the receiver then experience conversion in vacuum only, which leads to a broadband signal as the absence of disks in the path of the wave means that transmission is frequency-independent.

\begin{figure}
    \centering
    \includegraphics[width=\columnwidth]{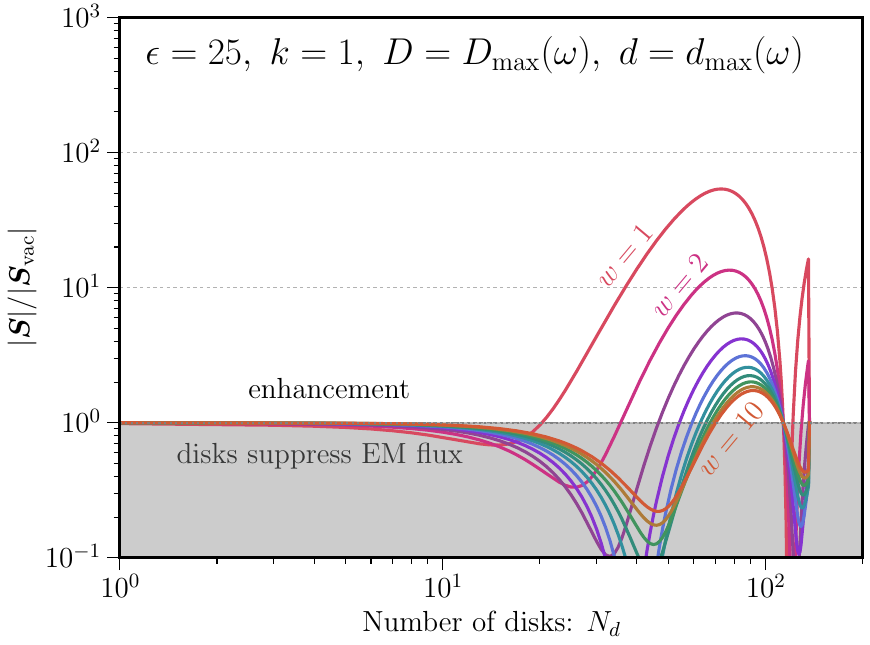}
    \caption{Performance of a dielectric haloscope in hybrid resonant/broadband mode. We plot Poynting vector ratios $|\vec{S}|/|\vec{S}_{\rm vac}|$ as a function of $N_d$, using $\ell = \SI{2}{m}$ for the length of the apparatus and fixing $k=1$ in the expression \eqref{eq:lmax} for $l_{\rm max}$. All disks are assumed to be placed on the side of the apparatus furthest from the receiver, so that its remaining length is vacuum. We show ten curves corresponding to $\w = 2\pi \, w \times \SI{e10}{Hz}$ with $w \in [1,10]$.}
    \label{fig:hybridNdisk}
\end{figure}

In the hybrid scenario, the optimal number of disks is frequency-dependent as shown in \cref{fig:hybridNdisk}, where we plot $|\vec{S}|/|\vec{S}_{\rm vac}|$ for ten different frequencies in the range $\w \in 2 \pi \times [10^{10},~10^{11}]\,\si{Hz}$. The largest boost is obtained for $\w = 2\pi \times \SI{e10}{Hz}$ and is about $|\vec{S}|/|\vec{S}_{\rm vac}| \sim 50$ for $N_d \sim 70$. For higher frequencies the boost shrinks and is achieved only upon the placement of more disks, until reaching $|\vec{S}|/|\vec{S}_{\rm vac}| \sim 2$ for $N_d \sim 90$ at $\w = 2\pi \times \SI{e11}{Hz}$.
This frequency dependence largely reflects the reduced amount of detector length reserved for the resonant operation at higher frequencies when fixing $k=1$, and is compensated by a larger fraction of the detector available for a broadband search. Alternatively, one could improve the resonant sensitivity at higher frequencies by increasing $k$ -- at the cost of reducing the corresponding broadband sensitivity.

This suggest that an optimal hybrid setup is a detector with $N_d = 73$ disks optimised for $\omega = 2\pi \times \SI{e10}{Hz}$, occupying about half of the $\ell = \SI{2}{m}$ apparatus. Such a setup optimises the resonant sensitivity to the lowest frequencies considered, while leaving around \SI{1}{m} of vacuum between the last disk and the receiver, over which GWs entering at an appropriate angle can be detected in broadband mode. More precisely, this is the case for GWs entering under an angle $|\theta| \gtrsim \arctan (2R/\ell_{\rm eff})$, where $R$ is the radius of the cylindrical apparatus and the quantity $\ell_{\rm eff}$ is obtained from the length of the vacuum section, $\ell_{\rm vac}$ and the wavelength of the GW, $\lambda = 1/f$ as
\begin{align}
    \ell_{\rm eff} = \frac{16 R^2 \ell_{\rm vac} - 2 R \lambda \sqrt{4\ell_{\rm vac}^2 + 16 R^2 - \lambda^2}}{16R^2-\lambda^2} \,.
    \label{eq:ell-eff}
\end{align}
This effective length arises from the requirement that the geometric optics limit used here is valid, i.e.\ that none of the GW wavefronts of length $\lambda$ pass through the disks on the way to the receiver. In the infinite frequency, $\ell_{\rm eff} = \ell_{\rm vac}$ as expected, while $\ell_{\rm eff} = 0$ for $\lambda \geq 2\ell_{\rm vac}$. For the configuration advocated above with $\ell_{\rm vac} \sim \SI{1}{m}$, the requirement on $\ell_{\rm eff}$ is such that the broadband experiment running concurrently with the resonant one can cover the frequency range $\w \gtrsim 2\pi \cdot \SI{1.5e8}{Hz}$ with a broadband readout.

For the hybrid setup the Poynting vector associated to disk-enhanced resonant GW-to-photon conversion is given by \cref{eq:Poynting} (neglecting the contribution of $E_p(\ell)$), while for conversion in vacuum for GW not passing through the disks before reaching the receiver, it is
\begin{align}
    \vec{S}_{\rm eff} \sim h^2 (\w\,\ell_{\rm eff}\,B_0)^2 \,.  
    \label{eq:PoyntingVacuum}
\end{align}
This equation is used for the broadband part of the hybrid setup we consider when discussing our sensitivity estimates in \cref{sec:sensitivity}.

Let us now consider fully broadband operation of the MADMAX device, which would imply removing the dielectric disks (thereby rendering the apparatus much less sensitive to its primary physics target, axions). As with the broadband part of the hybrid setup discussed above, the resulting Poynting vector is constructed entirely from the solution to Maxwell's equations in vacuum given, \cref{eq:3D_vac_p}. There is again an opening angle within which GW-to-photon conversion benefits from the full apparatus length $\ell$, given by $\theta_c \lesssim \arctan(2R/\ell)$. As long as the GW angle is less than this value, the flux density reaching the receiver scales the same way as \cref{eq:PoyntingVacuum}, only with $\ell$ on the RHS instead of $\ell_{\rm eff}$. This calculation matches the results in Ref.~\cite{Ejlli:2019bqj} which considered vacuum conversion of GWs to photons in light-shining-through-wall experiments.

If the experiment is operated in the fully broadband configuration, the high-frequency limit of the sensitivity band is determined by the highest frequency at which broadband readout can be achieved, which we expect to lie around ${\cal O}(10 - 100)$~GHz for MADMAX-like readout. At lower frequencies, if the regime $\w R \sim 1$ is reached, our analysis does not account for possible EM resonances in the apparatus, which could change the solution to Maxwell's equations in vacuum and is therefore not to be taken at face value. We plot this region in \cref{fig:sens} as a dashed green line. At even lower frequencies, $\w R \lesssim 1$, the walls of the MADMAX apparatus act as a waveguide, suppressing power flow to the receiver. For the dimensions of the MADMAX apparatus, this occurs for frequencies $\w/(2\pi) \lesssim \SI{3e8}{Hz}$, and is accounted for in our sensitivity estimates. 
In practice, we can model the response of the MADMAX apparatus in this regime as a cavity of resonant frequency $\w_c \sim 1/R$ being driven by a lower-frequency signal, leading to a suppression of the power by a factor $\w^4/(\w^2-\w_c^2)^2 \sim (\w R)^4$.

\section{Sensitivity estimates}
\label{sec:sensitivity}

\begin{figure*}
    \centering
    \includegraphics[width=\textwidth]{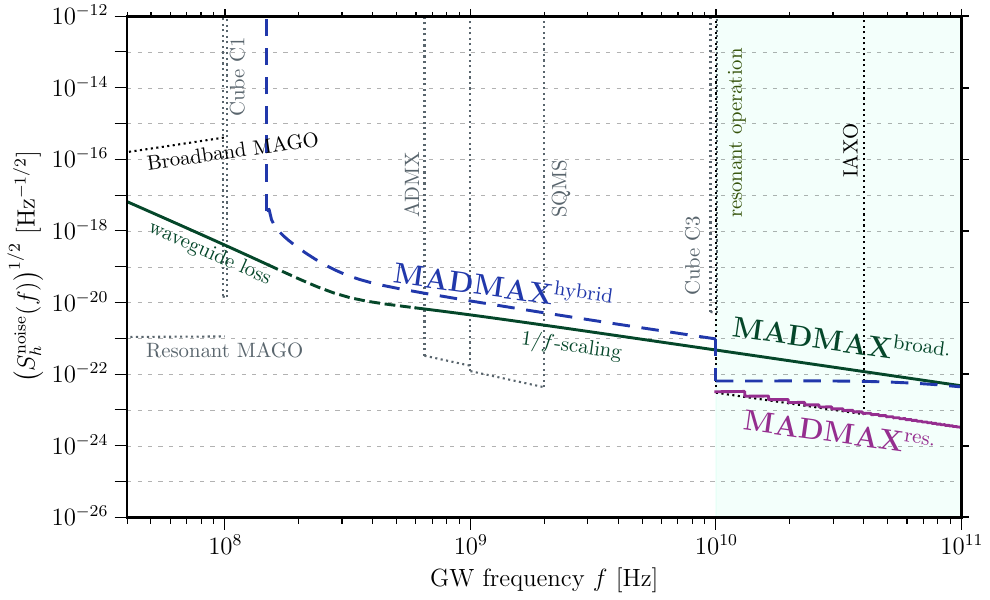}
    \caption{GW strain sensitivity for different operation modes of a dielectric haloscope with the dimensions of MADMAX, and for a GW incoming along the $x$-axis. (i) Broadband operation with \emph{no} dielectric disks (darker green). The dashed region around $\w R \sim 1$ indicates the regime where electromagnetic resonances of the apparatus can be relevant, implying that our results should be taken with a grain of salt. The label `$1/f$-scaling' indicates where the signal field grows linearly with $f$, resulting in a linear decrease in $S_h^{\rm noise}$. (ii) Fully resonant operation (purple) above \SI{10}{GHz}, assuming the optimal number of disks at each frequency, spaced such that the detector is maximally filled ($k = k_{\rm fill}$ in \cref{eq:lmax}). (iii) Hybrid resonant/broadband operation (dashed blue) with a stack of disks optimised for $f = \SI{1}{GHz}$ at the left end of the detector, choosing the smallest solution ($k=1$) for their spacing. For comparison, we show in dotted black projected limits of other proposals for broadband GW detectors 
    (MAGO 2.0~\cite{Berlin:2023grv} and low-frequency IAXO~\cite{Ringwald:2020ist})
    and in dotted this expression can gray proposals for resonant detectors 
    (resonant cavities~\cite{Berlin:2021txa, Navarro:2023eii}
    and MAGO 2.0~\cite{Berlin:2023grv}).
    }
    \label{fig:sens}
\end{figure*}

We are now ready to estimate the sensitivity of a MADMAX-like dielectric haloscope to the GW strain $h$. A useful figure of merit for this is the signal-to-noise ratio (SNR) of a hypothetical signal.
The optimal SNR can be written as
\begin{align}
    \text{SNR} \simeq \left(t_{\rm int}\int_0^\infty\,df\, \left(\frac{S_{\rm sig}(f)}{S_{\rm noise}(f)} \right)^2 \right)^{1/2} \ ,
\end{align}
where $S_{\rm sig,\,noise}$ are the signal and noise power spectral densities (PSDs), respectively,and $t_\text{int}$ is the integration time. The latter is given by the length of the data taking period for stochastic signals and by the duration of the signal for transient sources.
For a signal whose waveform is known and can be matched to the datastream in the time domain, a matched filtering analysis can be performed leading to an SNR with the same form, only with a ratio of signal and noise PSDs that is not squared. In both cases, it is instructive to rewrite the SNR in terms of the PSD of the GW $S_h(f)$, and the so-called ``noise-equivalent strain PSD'' $S_h^{\rm noise}(f)$. The former is defined for an approximately stationary process as
\begin{align}
    S_h(f) &\simeq \frac{1}{T} \langle \tilde{h}(f) \tilde{h}(f') \rangle \notag\\
           &=      \frac{1}{T} \left[ |\tilde{h}_+(f)|^2 + |\tilde{h}_\times(f)|^2 \right] \, ,
    \label{eq:Sh}
\end{align}
were $\langle\cdot\rangle$ denotes taking the ensemble average of Fourier transformed quantities $\tilde{h}(f)$, and $T$ is the integration time of the receiver. The noise-equivalent strain PSD is in turn defined as 
\begin{align}
    S_h^\text{noise}(f) &= \frac{S_h(f)}{S_\text{sig}(f)} S_\text{noise}(f) \,,
    \label{eq:Shnoise}
\end{align}
where $S_{\rm noise}(f)$ includes all noise contributions seen by the detector. For MADMAX, thermal noise is expected to dominate such that $S_{\rm noise}(f) \sim S_{\rm th}(f) = 2\,T_{\rm sys}$, where $T_\text{sys} = \SI{4.2}{K}$ is the system temperature of the apparatus. From these definitions, it is clear that we can rewrite the SNR in terms of the ratio $(S_h(f)/S_h^{\rm noise}(f))$. The noise-equivalent strain PSD is therefore a useful figure of merit that includes not only the intrinsic noise, but also the efficiency with which a general GW signal $S_h(f)$ is converted into a measured signal by the detector.

For the fully resonant operation mode, the signal power at the receiver, $P_\text{sig}(\omega) = \pi R^2 \vec{S}_x$ can be computed from \cref{eq:ErecR}. We then find for the PSD at the receiver
\begin{align}
    S^{\rm res}_\text{sig}(f) = {\cal C}^2 \pi R^2
                           \left(2 \pi f \ell B_0  N_d \right)^2 S_h(f) + {\cal O}(1/\sqrt{\epsilon})\,,
    \label{eq:sigPSDres}
\end{align}
where ${\cal C} \sim 0.4 \sqrt{\epsilon /25}$ denotes the numerical factor dropped in \cref{eq:ErecR}. Here $f = \omega/(2 \pi)$ denotes the GW frequency, which is identical to the frequency of the generated electromagnetic wave. For simplicity, here and in the following, we assume the GW to travel along the symmetry axis of the detector, $\theta = 0$. 
Plugging \cref{eq:sigPSDres} into \cref{eq:Sh}, we find for the strain-equivalent noise PSD in resonant mode
\begin{align}
    \left( S_h^\text{noise} \right)^{1/2}
        \! \sim \! \frac{10^{-22}/\si{Hz}^\frac{1}{2}}{{\cal C}/0.4} \!\!
              \left(\! \frac{\SI{1}{m^2}}{R \, \ell} \right) \!\!
              \left(\! \frac{\SI{10}{T}}{B_0} \right) \!\!
              \left(\! \frac{43}{N_d} \right) \!\!
              \left(\! \frac{10\,\si{GHz}}{f} \right) \!.
    \label{eq:sens-res}
\end{align}
Sensitivity estimates derived from this expression are shown as the purple curve in \cref{fig:sens}, where we have fixed $N_d = 43$, $\ell=\SI{2}{m}$, $\pi R^2 = \SI{1}{m^2}$, and have used $k = k_{\rm fill}$ to determine the disk spacing in \cref{eq:lmax}. The last assumption leads to the jaggedness of the sensitivity curve -- each of the steps in the curve corresponds to a change in $k_{\rm fill}$.

When the detector is operated in fully broadband mode without disks, the signal PSD is given by
\begin{align}
    S^{\rm broad}_\text{sig}(f) = \frac{1}{8}\pi R^2
                                  \left(2 \pi f \ell B_0 \right)^2 S_h(f) \ ,
    \label{eq:sigPSDbb}
\end{align}
with the same notation as above. This result holds again for a GW traveling along the symmetry axis of the detector, but the more general result for an arbitrary incoming angle can be directly computed from \cref{eq:3D_vac_p}. Sensitivity estimates derived from \cref{eq:sigPSDbb} following the same approach as above are shown as the solid green curve in \cref{fig:sens}. Across much of the frequency range of interest here, we observe sensitivity growth as $\propto 1/f$, corresponding to the $\propto \w \ell$ growth of the EM field amplitude at the receiver, see \cref{eq:3D_vac_p}. At small frequencies we include the effect of suppression of the flux arriving at the receiver due to the wavelength of the GW becoming larger than the radius of the apparatus. This implies that the boundaries of the detector act as a waveguide, suppressing the EM signal and thereby the sensitivity. Finally, we show the region around $\w R \sim 1$ as a dashed line to indicate that our calculation does not include various effects such as EM resonances associated to the detector size that could occur in this frequency range.

Finally, in the hybrid regime, the signal PSD for GWs received in broadband mode (i.e.\ not passing through the disks before hitting the receiver) is identical to \cref{eq:sigPSDbb} above, only with $\ell$ replaced by $\ell_{\rm eff}$ from \cref{eq:ell-eff}. Meanwhile the signal PSD for resonantly-enhanced signals exhibits a more complicated $N_d$ dependence than \cref{eq:sigPSDres}, the equivalent for fully resonant setups. For $N_d \lesssim 10$, the impact of the disks is small compared to conversion in vacuum, and the signal PSD is well-approximated by \cref{eq:sigPSDbb}. For larger number of disks, the scaling with $N_d$ is frequency-dependent, and can be inferred from the ratio $|\vec{S}|/|\vec{S}_{\rm vac}|$ shown in \cref{fig:hybridNdisk}. 
For the benchmark scenario with $N_d = 73$ disks we advocated for in \cref{sec:Hybrid}, the expected sensitivity is shown as a dashed blue line in \cref{fig:sens}. The choice $N_d = 73$ maximises the resonant boost at $f = \SI{e10}{Hz}$. For $f \geq \SI{e10}{Hz}$, the sensitivity is dominated by resonantly enhanced signals, and we show the result assuming a GW propagating along the detector axis $\theta = 0$. For $f \leq \SI{e10}{Hz}$, we compute the effective length travelled in vacuum as a function of frequency, and show the corresponding reduction in sensitivity. It is mostly a factor of 2 except at the lowest frequencies detectable.

Detection in the broadband regime has the advantage of being sensitive to short-duration signals that rapidly scan through signal frequency, such as the mergers of sub-solar mass black holes. These black holes must be of a primordial nature and could comprise some or all of the dark matter. As shown in Appendix~\ref{app:PBHdistance}, a detector with no disks is sufficient to detect mergers up to distances $d_{\rm PBH} \sim \SI{e-4}{pc}$ for chirp masses in the range $10^{-7}\,M_\odot \lesssim M_c \lesssim 10^{-5}\,M_\odot$.  In this context, we stress that the typically very short duration $t_\text{int} \sim 1/f$ and low phase coherence of the signal from merger events has been accounted for in the derivation of this estimate. Compared to e.g.\ GW searches using resonant cavities, which require sufficiently long and coherent signals to fully ring up the cavity, the induced electromagnetic waves in our broadband proposal simply track the wave front of the gravitational wave as it passes through the detector, inheriting the frequency spectrum and phase factor of the GW. This makes our proposal significantly more sensitive to these types of signals. However, to reach appreciable expected merger rates of order one per year (corresponding to a reach of $d_{\rm PBH} \gtrsim \SI{10}{kpc}$ assuming primordial black holes in aforementioned mass range make up a significant fraction of dark matter~\cite{Franciolini:2022htd,Domcke:2022rgu}) requires an implausibly large magnetised volume, since $d_{\rm PBH} \propto (B_0\,\ell\,R)$. We note that the required eight order of magnitude increase in the combination $(B_0\,\ell\,R)$ would also put cosmogenic sources of GW such as primordial phase transitions within reach.

\section{Conclusions}
\label{sec:conclusion}

We have analysed the prospects of GW detection in dielectric axion haloscopes, which consist of a stack of dielectric disks placed inside a magnetic field. We have focused specifically on the parameters of the planned MADMAX experiment primarily designed to search for axion dark matter. Similar to axions, GWs source propagating EM fields at the surfaces of the dielectric disks, which for a suitable geometry can interfere constructively to maximize the signal at the receiver. However, the relativistic nature of the GW leads to several important differences: (i)~unlike axion-to-photon conversion, GW-to-photon conversion occurs also in the vacuum between disks, enhancing the EM signal in the GW case. (ii)~The sourced EM waves inherit the position dependent GW phase, leading to new conditions to ensure resonant operation, including in particular a requirement on the effective disk thickness $d = \sqrt{\epsilon} \cos\theta_\epsilon$, and limiting the maximal number of disks that can be used to enhance the signal. (iii)~The sourced EM waves inherit the propagation direction of the GW, reducing the manufacturing requirements on the disk surface. Consequently, this setup provides an example where the parametric sensitivity to GWs cannot be obtained by a simple dimensional-analysis recasting of the axion sensitivity.

Compared to previous proposals based on the GW-to-photon conversion in vacuum in the presence of a strong magnetic field, we find that placing dielectric disks in a magnetised apparatus can lead to a slight improvement in the noise-equivalent strain sensitivity, by up to a factor of $10$. However, this improvement comes at the cost of making the apparatus a resonant detector that is maximally sensitive only in an extremely narrow range of GW frequencies, while suppressing the signal at other frequencies (see \cref{fig:resonances} in \cref{app:transfermatrix}). The enhancement at the resonance frequency moreover relies on exquisite manufacture quality and placement of the dielectric disks.

We therefore also consider two alternative approaches to use the MADMAX apparatus, namely hybrid resonant/broadband and fully broadband operation. In the latter, the disks are removed entirely, and we find a slightly reduced sensitivity of at best $S_h^{1/2} \sim \SI{e-21}{Hz^{-1/2}} \times (\SI{10}{GHz} / f)$, which is nevertheless competitive with other recent proposals. Broadband operation has the advantage of being able to cover a wide range of possible signal frequencies without requiring any tuning. The hybrid approach combines resonant and broadband operation, with a fixed number of disks placed in one half of the detector, leaving the other half as vacuum. GWs at frequencies above $f \gtrsim \SI{e10}{Hz}$ then benefit from resonant enhancement of the sensitivity (with the same requirements on disk placement and thickness as in the fully resonant setup), while GW frequencies between $f \in [0.015,1] \times \SI{e10}{Hz}$ can be detected in broadband mode, albeit with a reduced opening angle and therefore effective detector length.

Detectors designed to search for axion dark matter can often be re-purposed to search for high frequency gravitational waves (see e.g.~\cite{Berlin:2021txa,Domcke:2022rgu,Domcke:2023bat,Domcke:2024mfu} for other work along these lines). However, the wide range of possible gravitational wave signals, covering various dimensions such as central frequency, duration and coherence make designing the ideal detector a complicated task, with each detector having its advantages and disadvantages. Dielectric haloscopes such as MADMAX have the advantage of being operable in multiple regimes, each of which can benefit from different experimental aspects, making them an attractive option in the enduring search for high-frequency gravitational waves.

\vspace{5mm}
\textit{Acknowledgements.} We thank Camilo Garcia-Cely, Sung Mook Lee, Bela Majorovits, Jamie McDonald, and Pascal Pralavorio for useful comments and discussions, and G\'eraldine Servant for encouragement. We are particularly grateful to Jan Sch\"utte-Engel for his questions on the comparison to the relativistic axion case, which lead to crucial insights on the resonance conditions. We thank the MADMAX collaboration for the invitation to their collaboration meeting at CERN and various helpful discussions. The work of SARE was supported by SNF Ambizione grant PZ00P2\_193322, \textit{New frontiers from sub-eV to super-TeV}.

\appendix
\onecolumngrid
\interfootnotelinepenalty=10000 

\section{Conventions}
\label{app:conventions}

We list here the various conventions we use in the derivation of our results. For the gravitational wave (GW), we work in the transverse traceless frame where the metric perturbation is written as
\begin{align}
    h_{ij}^{\rm TT} = \left(h^+ \hat{e}^+_{ij} + h^\times \hat{e}^\times_{ij} \right) e^{-i \w(t-\vec{\hat{k}}\cdot \vec{x})} \,, 
\end{align}
with the explicit form of the polarization tensors given by
\begin{align}
 \hat e_{ij}^+ = u_i u_j - v_i v_j \,, \quad \hat e_{ij}^\times = u_i v_j + v_i u_j \,,
\end{align}
where $\vec{\hat v} = (- s_\phi, c_\phi,0)$, $\vec{\hat u} = \vec{\hat v} \times \vec{\hat k}$ and the normalized GW wave vector $\vec{\hat k}$ form an orthonormal system. The polarization tensors are transverse  $k_i \hat e^{ij}_\lambda = 0$ and orthogonal $\hat{e}^\lambda_{ij}\hat{e}^{\lambda',ji} = 2 \delta^{\lambda\lambda'}$.
All electromagnetic fields generated by the passing GW inherit the time-variation and propagation phasor, such that when evaluating observables, we must take the real part of the relevant quantities. This results in the factor $(1/2)$ and the $\text{Re}[\ldots]$ appearing in our definition for the Poynting vector in Eq.~\eqref{eq:Poynting}.

For the power spectral densities, we use the convention that the two-point correlation function of a complex quantity $x(f)$ in frequency-space is related to the two-sided PSD as
\begin{align}
    \langle x(f) x^*(f')\rangle \equiv S_x(f) \delta(f-f') \ .
\end{align}
The PSD is in turn related to the auto-correlation function of the quantity $x(t)$ in the time domain as
\begin{align}
    A_x(\tau'-\tau) = \langle x(\tau)x^*(\tau') \rangle = \int_{-\infty}^\infty df\,e^{-2\pi i f(\tau'-\tau)} S_x(f) \ ,
\end{align}
where the minus sign in the exponent reflects our phase (and therefore Fourier transform) convention.

\section{The Transverse-Traceless Frame in the High-frequency Limit}
\label{app:TT-HFlimit}

In the main body of this paper, we work in the Transverse-Traceless (TT) frame to calculate the effect of the passing GW on the EM fields. More specifically, we treat the background magnetic field as being static in the TT frame, which amounts to assuming that it is freely falling. In this section, we justify this approach.

The key to our argument is the hierarchy between the wavelength $\lambda$ of the GW and the dimensions of a typical dielectric haloscope. For GW frequencies in the range $\w/(2\pi) \in (10^8,~10^{11})\,\text{Hz}$, the former is $\lambda \in (3\times10^{-3},~3)\,\text{m}$. These wavelengths should be compared with the typical size of the apparatus, which is a few meters.\footnote{Corresponding to characteristic frequencies $\w_{\rm det} \lesssim 2\pi v\times10^{8}\,\text{Hz}$, where $v=c~(c_s \ll1)$ for EM (elastic) waves.} As a result, we work almost exclusively in the $\w \ell \gg 1$ regime where the usual long-wavelength approximation to the proper detector frame breaks down. An option would be to use the all-orders in $\w \ell$ formulation of the proper detector frame~\cite{Berlin:2021txa}, but this rapidly complicates the computation. Instead, it is well known that the TT frame offers a valid description of the interaction of GWs with objects when $\w \ell \gg 1$, as they are all effectively free-falling in this regime (see e.g.\ discussions in \cite{Bringmann:2023gba, Ratzinger:2024spd}). Let us make this argument more explicit and see in particular how it applies to the applied magnetic field.

To understand how objects in the laboratory frame behave in the large $\w \ell$ limit, we start by recalling the definition of the geodesic deviation equation, which is derived by considering two geodesics separated by a small distance $\xi$, each respecting their own geodesic equations
\begin{align}
    &\frac{d^2x^\mu}{d\tau^2} + \Gamma^{\mu}_{\nu\rho}(x) \frac{dx^\nu}{d\tau} \frac{dx^\rho}{d\tau} = 0 \ , \\
    &\frac{d^2(x+\xi)^\mu}{d\tau^2} + \Gamma^{\mu}_{\nu\rho}(x+\xi) \frac{d(x+\xi)^\nu}{d\tau} \frac{d(x+\xi)^\rho}{d\tau} =0\ .
\end{align}
Evaluating the difference between these equations at a point $x = 0$, and assuming that the detector is at rest or moving non-relativistically, such that only $d\xi^0/d\tau \neq 0$, we see that all we need to compute is $\Gamma^{\mu}_{00}(\xi)$. A subtlety is that this Christoffel symbol should be evaluated using the metric in the laboratory frame that includes terms to all orders in $\w \vec{\hat{k}}\cdot \vec{x}$.  This metric was worked out in Refs.~\cite{Marzlin:1994ia,Rakhmanov:2014noa}, and resummed in Ref.~\cite{Berlin:2021txa} for a plane wave. For computing $\Gamma^{\mu}_{00}(\xi)$, we need the all-orders metric components $h_{00}$ and $h_{0i}$ given by~\cite{Marzlin:1994ia,Rakhmanov:2014noa,Berlin:2021txa,Domcke:2022rgu}
\begin{align}
    \nonumber h_{00} &\equiv -2\left(\sum_{n=0}^\infty \frac{n+3}{(n+3)!}\left( R_{k0l0}\right)^{(n)} \xi^k\xi^l \xi^{j_n}\right) \\
    &= -\w^2 h_{kl}^{\rm TT}(x) \xi^k \xi^l \left(-\frac{i}{\w \vec{\hat{k}}\cdot \vec{x}} + \frac{1-e^{-i\w \vec{\hat{k}}\cdot \vec{x}}}{(\w \vec{\hat{k}}\cdot \vec{x})^2}\right) \ , 
    \label{eq:h00}
    \\
    \nonumber h_{0k} &\equiv -2\left( \sum_{n=0}^\infty\frac{n+2}{(n+3)!} \left( {R^k}_{j_1 0 j_2}\right)^{(n)} \xi^{j_1} \xi^{j_2} \xi^{j_{n}}\right) \\
    &=-\w^2 \left(h_{kl}^{\rm TT}(x) \xi^l (\vec{\hat{k}}\cdot \vec{x}) - \hat{k}_k h_{jl}^{\rm TT}(x) \xi^j \xi^l \right) \left(-\frac{i}{2\w \vec{\hat{k}}\cdot \vec{x}} - \frac{e^{-i\w \vec{\hat{k}}\cdot \vec{x}}}{(\w \vec{\hat{k}}\cdot \vec{x})^2} -\frac{1 - e^{-i\w \vec{\hat{k}}\cdot \vec{x}}}{(\w \vec{\hat{k}}\cdot \vec{x})^3}\right) \ .
    \label{eq:h0i}
\end{align}
As a result, we can write the geodesic deviation equation \emph{to all orders} as
\begin{align}
    \frac{d^2 \xi^k}{d\tau^2} + \left(\pd_t h_{0k} - \frac{1}{2}\pd_k h_{00} \right)\frac{dx^0}{d\tau} \frac{dx^0}{d\tau} =0 \ .
\end{align}
Note that we could have easily obtained this expression from the geodesic deviation equation without expanding in small $\xi$, but instead using the definition of the Christoffel symbols and our closed-form expressions for the metric.

We can easily check that in the $\w \ell \ll 1$ limit, this recovers the usual expression
\begin{align}
    \ddot{\xi}_k = \frac{1}{2}\ddot{h}_{kl}^{\rm TT} \xi^l \ 
\end{align}
whereby the effect of a GW is that of a Newtonian force. 
To understand the impact of this force on the external magnetic field (i.e.\ on the experimental apparatus generating the magnetic field), we need to compare this force to the internal mechanical forces in those materials. Internal forces act as $\nabla^2 \vec{\xi},~\nabla(\nabla \cdot \vec{\xi})$, and therefore scale as $c_s^2\xi/\ell^2$, where $c_s\ll 1$ is the speed of sound. For $\w \ell \ll c_s$, the GW force is weaker than internal forces and its effect is therefore suppressed. For $c_s \ll \w \ell \ll 1$, the long wavelength limit still applies, but the GW force overcomes internal forces and dominates the equation of motion for $\xi$.

Of greater interest to our situation is the opposite limit, where $\w \ell \gg 1$, where explicit computation shows that the geodesic deviation equation becomes
\begin{align}
    \ddot{\xi}_k = \frac{1}{2}\ddot{h}_{kl}^{\rm TT}  \xi^l  + \mathcal{O}(1/\w \vec{\hat{k}}\cdot \vec{x}) \ ,
    \label{eq:geodesic_HF}
\end{align}
as long as the transverse directions are smaller than the direction along $\vec{\hat{k}}\cdot\vec{x}$.
Since $\omega \ell \gg 1$ also means $\w \ell \gg c_s$, on the basis of the discussion above, all internal forces in the material can be safely neglected.
The coupling of the GW to higher phonon modes is suppressed by the missmatch between the dispersion relations of GWs and phonons in the material, see e.g.~Ref.~\cite{Berlin:2023grv} for an explicit example.
As a result, it is legitimate to treat the apparatus generating the magnetic field, and therefore the magnetic field itself, as being free-falling in the high-frequency limit.
Since the coordinates in the TT frame are set by free-falling test-masses (which move in the proper detector frame according to \cref{eq:geodesic_HF}), the apparatus and consequently the magnetic field appear static in the TT frame.

Finally, in solving Maxwell's equations including the effective current due to the presence of the GW, we found that the particular solution grew as $\w \ell$. As discussed in the main text, this is a result of the mass degeneracy between the GW and photon. However, one might wonder how to see that this result holds also in the laboratory frame. As we see in~\cref{eq:h00} and \cref{eq:h0i}, the laboratory frame metric to leading order in $\w\ell \ll 1$ scales as $h\sim (\w \ell)^2 h^{\rm TT}$. However, much of the parameter space we are interested in lies in the opposite limit, $\w\ell \gg 1$. In that limit, explicit computation shows that both $h_{00}$ and $h_{0i}$ scale as $h \sim (\w \ell) h^{\rm TT}$. Therefore, had we opted to obtain the particular solution to Maxwell's equations in the laboratory frame, we would have obtained the same result as in the TT frame, as expected.

\section{Additional Details on Transfer Matrix Formalism}
\label{app:transfermatrix}

In this appendix, we expand on the discussion in \cref{sec:TransferMatrix} by deriving the explicit expressions for all the ingredients to describe the multi-disk system using the transfer matrix formalism. Where applicable, we follow the procedure and notation developed for axions in Ref.~\cite{Millar:2016cjp}. We will discuss in particular the conditions on disk width and spacing required to optimize the sensitivity to GWs.

As in the main text, we consider a setup with $2 N_d + 1$ regions, alternating between vacuum regions of length $D$ and in-medium regions of length $d$ with a dielectric constant $\epsilon > 1$. We consider a cylindrical setup of disks stacked along $x$-axis, with constant magnetic field pointing in $z$-direction and EM waves propagating in the $x$--$z$ plane. In each region we consider left- and right-moving EM waves, with the 2-component vector $E_n = (R_n, L_n)$ denoting their amplitude on the left edge of the $n$th region. Throughout our expressions, we will drop a global factor $\exp(- i \omega (t - s_\theta z) )$, common to the GW and all induced EM fields.
The propagation matrices describing the phase evolution across these regions are given by
\begin{align}
    P_v = \text{diag}(e^{i \omega D c_\theta}, e^{-i \omega D c_\theta}) \,, \quad P_m = \text{diag}(e^{i \omega \sqrt{\epsilon} d c_{\theta_\epsilon}}, e^{-i \omega \sqrt{\epsilon} d c_{\theta_\epsilon}})\,,
    \label{eq:prop}
\end{align}
for the vacuum and disk regions, respectively. Here $\theta$ is the angle between the right-moving EM wave and the $x$-axis, $c_\theta \equiv \cos\theta$, and $\theta_\epsilon$ is the angle of the refracted right-moving wave in medium, defined by the relation $\sqrt{\epsilon} \sin \theta_\epsilon = \sin \theta$. When the EM waves cross a disk surface, they obey the usual EM boundary condition at the interface of a medium,
\begin{align}
    \vec{B}_m = \vec{B}_v\,, \quad \vec{E}^\parallel_m = \vec{E}^\parallel_v\,, \quad \epsilon \vec{E}^\perp_m = \vec{E}^\perp_v \,,
    \label{eq:boundaryC}
\end{align}
where $\parallel$ and $\perp$ refer to the contributions parallel and perpendicular to the interface, respectively. These boundary conditions distinguish $s$-polarized EM waves (orthogonal to the plane of incidence) and $p$-polarized waves (perpendicular to the plane of incidence). At a vacuum-to-medium interface this is encoded in the matrices
\begin{align}
  G^p_{mv} = \frac{1}{2} \begin{pmatrix} \frac{1}{\sqrt{\epsilon}} + \frac{c_\theta}{c_{\theta_\epsilon}} & - \frac{1}{\sqrt{\epsilon}} + \frac{c_\theta}{c_{\theta_\epsilon}} \\
  - \frac{1}{\sqrt{\epsilon}} + \frac{c_\theta}{c_{\theta_\epsilon}} & \frac{1}{\sqrt{\epsilon}} + \frac{c_\theta}{c_{\theta_\epsilon}} \end{pmatrix} \,,
  \qquad
   G^s_{mv} = \frac{1}{2 \sqrt{\epsilon}} \begin{pmatrix} \sqrt{\epsilon} + \frac{c_\theta}{c_{\theta_\epsilon}} & \sqrt{\epsilon} -  \frac{c_\theta}{c_{\theta_\epsilon}} \\
  \sqrt{\epsilon} -  \frac{c_\theta}{c_{\theta_\epsilon}} & \sqrt{\epsilon} + \frac{c_\theta}{c_{\theta_\epsilon}} \end{pmatrix} \,,
  \label{eq:Gmv}
\end{align}
whereas the inverse process, a medium-to-vacuum boundary yields
\begin{align}
 G^p_{vm} = \frac{1}{2} \begin{pmatrix} \sqrt{\epsilon} + \frac{c_\theta}{c_{\theta_\epsilon}} & - \sqrt{\epsilon} +  \frac{c_\theta}{c_{\theta_\epsilon}} \\
  - \sqrt{\epsilon} +   \frac{c_\theta}{c_{\theta_\epsilon}} & \sqrt{\epsilon} + \frac{c_\theta}{c_{\theta_\epsilon}} \end{pmatrix} \,,
  \qquad
 G^s_{vm} = \begin{pmatrix} 1 + \sqrt{\epsilon} \frac{c_\theta}{c_{\theta_\epsilon}} & 1 - \sqrt{\epsilon} \frac{c_\theta}{c_{\theta_\epsilon}} \\
 1 - \sqrt{\epsilon}   \frac{c_\theta}{c_{\theta_\epsilon}} & 1 + \sqrt{\epsilon} \frac{c_\theta}{c_{\theta_\epsilon}} \end{pmatrix} \,.
  \label{eq:Gvm}
\end{align}
These expressions allow to explicitly construct the effective single-segment transfer matrix $T_1 = G_{vm} P_m G_{mv} P_v$ introduced in the main text.
Note that so far, this is simple geometric optics for a multi-disk system, independent of the presence of any axions or GWs.

With GWs included, we need to include an effective source term in the inhomogeneous Maxwell equation. It can easily be verified that a particular solution of this equation is given by \cref{eq:3D_vac_p} in vacuum and by \cref{eq:3D_med_p} in medium. These particular solutions are not unique as one can always add plane wave solutions (solutions to the source-free equation). The solutions only become unique once appropriate boundary conditions are imposed. The particular solution in vacuum given in \cref{eq:3D_vac_p} has been chosen so as to give a vanishing plane wave contribution at $x=0$, i.e.\ no right-moving plane waves at the left edge of the detector.\footnote{
The remaining part of this particular solution can be understood as the EM wave sourced at the boundary of the external magnetic field. Approximating the boundary at $x \simeq 0$ by $\vec{B}(\vec{x}) = \Theta(x) B_0 \hat{\vec{e}}_z$ gives rise to an effective surface current $\vec j_\text{eff}^\text{surf} = (\hat{\vec{e}}_x \times M_\text{eff})\delta(x) = B_0 \delta(xL) e^{- i \omega (t - \hat{\vec{k}} \cdot \vec{x})} (0, h_+ c_\theta, h_\times)$~\cite{Domcke:2023bat}. Matching boundary conditions at this interface gives rise to an EM wave in the magnetic field region.
}
The particular solution in medium \cref{eq:3D_med_p} has been chosen to take a conveniently simple form. By construction, our results do not change if we choose a different particular solution to Maxwell's equations, such as a sawtooth-shaped function which is given by \cref{eq:3D_med_p} inside the disks, vanishes just beyond the right edge of each disk, and then grows linearly up to the left edge of the next disk. The motivation for such a choice would be that it ensures $\vec{E}_v^p$ is the same in all vacuum regions.

In the transfer matrix formalism, the EM fields induced by the GW enter through the source terms $S_{mv}$ and $S_{vm}$ at the vacuum-to-medium and medium-to-vacuum interface, respectively. They are obtained by solving the inhomogeneous Maxwell equations at these interfaces. Since we have set up the transfer matrix formalism to propagate the solutions for the EM waves from left to right, we determine the remaining free coefficients in the source terms (corresponding to adding plane wave solutions to the particular solutions \cref{eq:3D_med_p} and \cref{eq:3D_vac_p}) by imposing a vanishing amplitude for all the plane waves sourced to the left side of the respective interface, i.e.\ the source term is set up to generate left- and right-moving EM fields only to the right of the respective interface. 
For a GW with incident angle $\theta$ in the $x$--$z$ plane, allowing for right(left)-moving plane waves with angle $\theta$ ($-\theta$) in vacuum, and correspondingly $\theta_\epsilon$ ($-\theta_\epsilon$) inside the disks, provides solutions to \cref{eq:boundaryC}. The source terms for the vacuum-to-medium interface of the $n$th disk are then obtained as
\begin{align}
    S_{mv}^p &= \frac{B_0 h^\times e^{i x_n \omega c_\theta}}{8 \sqrt{\epsilon} (\epsilon - 1) c_{\theta_\epsilon}}
    \begin{pmatrix}
      \sqrt{\epsilon } (-2 i x_n \omega  (\epsilon -1) c_\theta
                        + (\epsilon -3) c_{2\theta} - \epsilon - 1)
                      -2 c_{\theta_\epsilon} [(\epsilon+1) c_\theta
                                            + i x_n \omega (\epsilon-1)] \\[0.1cm]
      2 c_{\theta_\epsilon} [(\epsilon+1) c_\theta + i x_n \omega (\epsilon-1)]
      +\sqrt{\epsilon } [-2 i x_n \omega (\epsilon-1) c_\theta
                         + (\epsilon-3) c_{2\theta} - \epsilon - 1]
    \end{pmatrix} \,,
    \label{eq:Spmv} \\
    S_{mv}^s &= \frac{B_0 h^+ e^{i x_n \omega c_\theta}}{4 \sqrt{\epsilon} (\epsilon - 1) c_{\theta_\epsilon}}
    \begin{pmatrix}
      -c_{2\theta}-i \left(x_n \omega  (\epsilon -1) c_{\theta}+\sqrt{\epsilon } c_{\theta_\epsilon} (x_n \omega  (\epsilon -1)-2 i c_{\theta})\right)-\epsilon \\[0.1cm]
      c_{2\theta}+i \left(x_n \omega  (\epsilon -1) c_{\theta}+\sqrt{\epsilon } c_{\theta_\epsilon} (-x_n \omega  (\epsilon -1)+2 i c_{\theta})\right)+\epsilon
   \end{pmatrix} \,,
   \label{eq:Ssmv}
\end{align}
and the corresponding terms for the medium-to-vacuum interface are given by
\begin{align}
    S_{vm}^p &= \frac{B_0 h^\times e^{i (x_n + d) \omega c_\theta}}{4 c_\theta}
                \begin{pmatrix}
                    1 + 2 i (x_n + d) \omega c_\theta - 4 c_\theta^2 /(\epsilon - 1) \\
                   -c_{2 \theta}
                \end{pmatrix}\,,
    \label{eq:Spvm} \\
    S_{vm}^s &= \frac{B_0 h^+ e^{i (x_n + d) \omega c_\theta}}{4 c_\theta}
                \begin{pmatrix}
                    1 + 2 i (x_n + d) \omega c_\theta - 4 c_\theta^2 /(\epsilon - 1) \\
                   -1
                \end{pmatrix} \,,
    \label{eq:Ssvm}
\end{align}
with $x_n = D n + (n - 1) d$ indicating the left edge of the $n$th disk. 
For a GW incoming along the symmetry axis of the detector, $\theta = \theta_\epsilon = 0$, and the above expressions reduce to Eqs.~\eqref{eq:sourceVacMed} and \eqref{eq:sourceMedVac} in the main text. 

\begin{figure}
    \centering
    \includegraphics[width=0.6\textwidth]{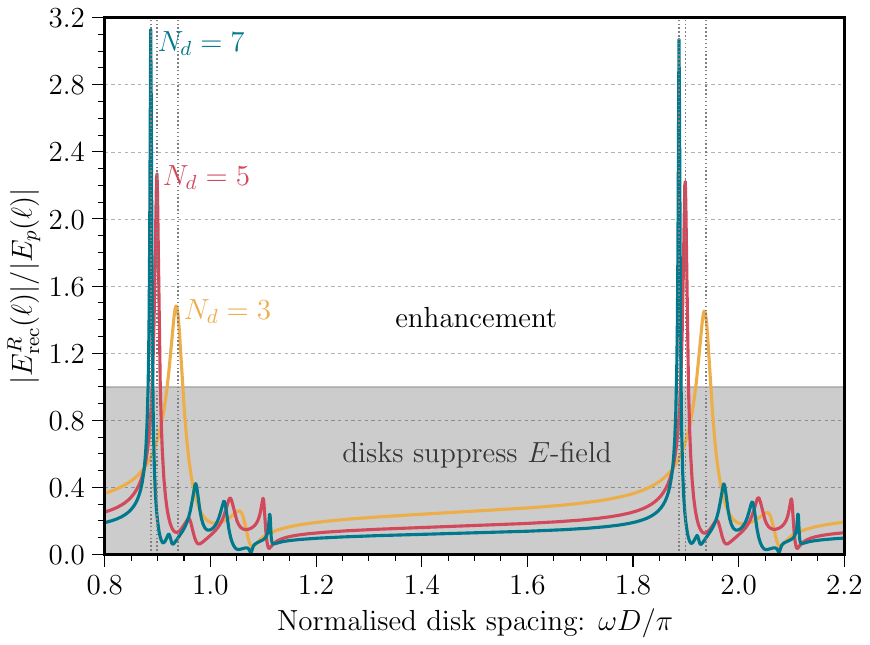}
    \caption{The ratio of the electric field at the rightmost edge of a detector with $N_d$ disks relative to the particular solution, $|E_{\rm rec}^R(\ell)|/|E_p(\ell)|$ as a function of the normalised disk spacing. We show results for $N_d=3,~5,~7$ in yellow, red and blue respectively. Note that the resonances where the field value is enhanced correspond almost exactly to the analytic values of $D_{\rm max}$, shown in the figure as dotted vertical lines. The agreement improves with larger $N_d$, as noted in the main text.}
    \label{fig:resonances}
\end{figure}

From \cref{eq:prop,eq:Gmv,eq:Gvm,eq:Spmv,eq:Ssmv,eq:Spvm,eq:Ssvm}, we construct the effective single disc source term $M^{(n)} = S_{vm} + G_{vm} P_m S_{mv}$ and the total effective source term
\begin{align}
     M_\text{tot} = \sum_{n = 1}^{N_d} T_1^{N_d-n} M^{(n)}
\end{align}
appearing in the expression~\eqref{eq:ErecE0}. $M_\text{tot}$ relates the EM fields at the two ends of the detector system.  When evaluated for the analytic optimal disk thickness $d_{\rm max}$ from \cref{eq:dmax}, and disk separation $D_{\rm max}$ from \cref{eq:lmax}, the two components of $M_\text{tot}$ can be approximated in the large $\epsilon$, small $N_d$ limit as
\begin{align}
   \left| [M_\text{tot}]_{R(L)} \right| \simeq \frac{1}{32}\pi^3 N_d (\w \ell) h B_0 \sqrt{\epsilon} \ ,
   \label{eq:Mtot_est}
\end{align}
where we have taken $\theta = 0$. We observe that at leading order in the large $\epsilon$, small $N_d$ limit, the two components of $M_\text{tot}$ are identical, supporting a posteriori the need for a vanishing reflection coefficient $\mathsf{r}_R$ to avoid any cancellation between the two terms in \cref{eq:ErecDet}. We show in  \cref{fig:resonances} the remarkable agreement between the analytically obtained ideal disc separation $D_\text{max}$, see \cref{eq:lmax}, and peaks in the signal strength obtained using the full transfer matrix formalism developed here.

\begin{figure}
    \centering
    \includegraphics[width=0.6\textwidth]{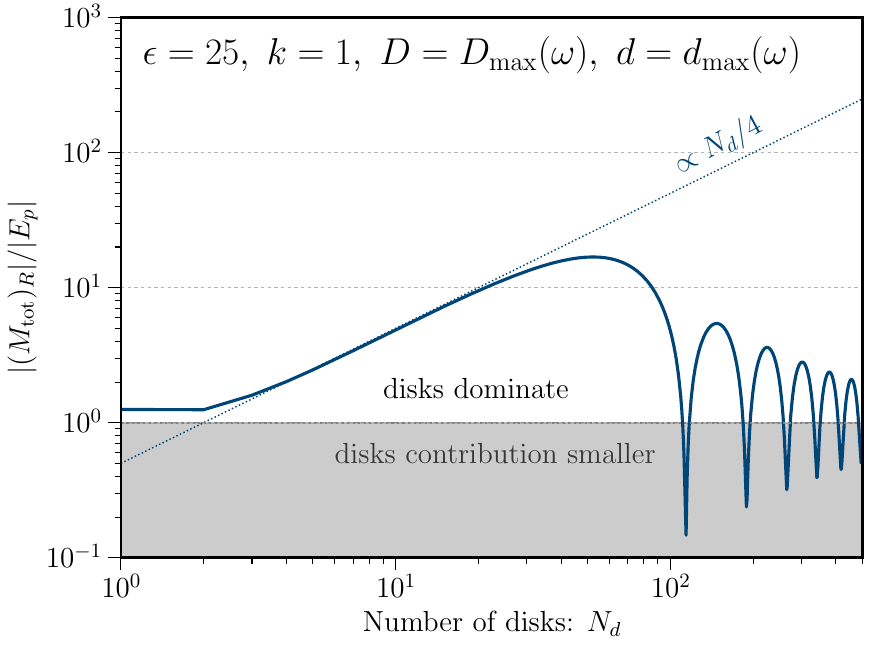}
    \caption{The ratio of the disk contribution to the full electric field relative to the particular solution, $|(M_{\rm tot})_R|/|E_p|$ as a function of the number of disks $N_d$. We have fixed the disk thickness and separation to their analytic optimum values, $d_{\rm max}$ and $D_{\rm max}$, respectively. We observe $\propto N_d$ growth up to moderate $N_d$, at which point the disk contribution begins to shrink relative to the vacuum contribution.}
    \label{fig:Nd-scaling}
\end{figure}

The parametric scaling of \cref{eq:Mtot_est}, and consequently the parametric estimate in \cref{eq:ErecR} in the main text, is confirmed numerically in \cref{fig:Nd-scaling}, which initially shows a linear growth of $(M_\text{tot})_R$ with the number of disks. The linear growth with the detector length $\ell$ is common to both $(M_\text{tot})_R$ and $E_p$, and hence drops out of the depicted ratio.
As we continue to increase the number of disks, destructive interference between different subsystems, i.e.\ within the different terms in $[M_\text{tot}]_R$, becomes a limiting factor. While the condition~\eqref{eq:lmax} ensures that the total $N_d$-disk system is maximally transmissive, subsystems with $n < N_d$ disks (as seem by EM waves sourced at all but the first disk) can feature significant reflection. At large $N_d$, this eventually limits the enhancement with $N_d$ and results in an oscillatory behaviour of the boost factor as a function of the number of disks. Numerically, we find that this happens around $N_d \gtrsim 50 \sqrt{\epsilon/25}$, see \cref{fig:Nd-scaling}. We note that this observation restricts only the useful number of disks, but not the disk spacing and detector length. For fixed number of disks, the sensitivity can still be improved by increasing $k$ in \cref{eq:lmax}, and thereby $\ell$.

\section{Emitted EM waves for a Fully Arbitrary Incoming GW}
\label{app:2angles}

We present here the particular solutions to Maxwell's equations in the presence of a GW traveling in an arbitrary direction with respect to the detector axis (as opposed to GW traveling in the $x$--$z$ plane considered so far). These are the input required to generalize the transfer matrix formalism to arbitrary 3D systems. We refrain from listing the full expressions for $P_{v,m}$, $G_{mv,vm}^{s,p}$, and $S_{mv,vm}^{s,p}$ in the 3D case here as they are rather cumbersome.

In the medium, a particular solution is given by
\begin{align}
    \vec{E}_m(\theta,\phi) &= \frac{B_0 e^{-i \w(t-\vec{\hat{k}}\cdot\vec{x})}}{\epsilon-1}
        \begin{pmatrix}
           -c_\theta \left(h_+ s_\phi + h_\times c_\phi s_\theta \right) \\
            c_\theta \left(h_+ c_\phi - h_\times s_\phi s_\theta \right) \\
            h_\times c^2_\theta
        \end{pmatrix} \,.
\end{align}
Meanwhile in vacuum, a particular solution is
\begin{align}
    \vec{E}_v(\theta,\phi) &= \frac{B_0 e^{-i \w(t-\vec{\hat{k}}\cdot\vec{x})}}{2}
        \begin{pmatrix}
            i \w x \left(h_\times s_\theta + h_+ t_\phi \right) \\[0.1cm]
           - i \w x\left( h_+ - h_\times s_\theta t_\phi \right)  \\
           -h_\times - h_+\frac{t_\phi}{s_\theta}  -i \w x h_\times \frac{c_\theta}{c_\phi}
        \end{pmatrix} \,.
\end{align}
It should be borne in mind that as particular solutions to Maxwell's equations, these are not unique, and are subject to the imposition of boundary conditions. In both the vacuum and medium particular solutions above, the propagation vector is $\vec{\hat{k}} = (c_\phi c_\theta,  s_\phi c_\theta, s_\theta)$.

\section{Sensitivity to Primordial Black Hole Binaries}
\label{app:PBHdistance}

Given the frequencies at which dielectric haloscopes are most sensitive, one of the most interesting signals they could detect arises from mergers of sub-solar mass black holes. These would almost certainly be of primordial origin, since generic stellar stability arguments imply that black holes formed in the late Universe should all have masses $m \gtrsim M_\odot$.

Black hole binaries emit coherent GWs at a frequency $f(t)$ which evolves with time as the binary radius shrinks in time. The strain can be approximately modelled as~\cite{Maggiore:2007ulw}
\begin{align}
    h(t) \simeq \frac{4}{d} \left( G M_c \right)^{5/3}\left( \pi f(t)\right)^{2/3} \cos\left( f(t) + \p \right) g(\theta,\phi) \,,
    \label{eq:pbh-h-of-t}
\end{align}
where $M_c$ is the chirp mass of the binary. For equal-mass black holes, it is given by $M_c = m/2^{1/5}$. Fourier-transforming, \cref{eq:pbh-h-of-t} becomes~\cite{Maggiore:2007ulw}
\begin{align}
    \tilde{h}(f) \simeq \left(\frac{5}{6}\right)^{1/2} \frac{1}{2\pi^{2/3}} \frac{1}{d} \left( G M_c \right)^{5/6} f^{-7/6} g(\theta,\phi) \,.
    \label{eq:pbh-h-of-f}
\end{align}
The function $g(\theta,\phi)$ captures the dependence on sky position of the signal with respect to the detector. The average of this quantity is typically $\mathcal{O}(1)$, so in what follows we will assume $|g(\theta,\phi)|^2 = 1$ for simplicity. Note that $h(t)$ and $h(f)$ have an explicit dependence on the distance to the binary $d$.

\begin{figure}
    \centering
    \includegraphics[scale=0.9]{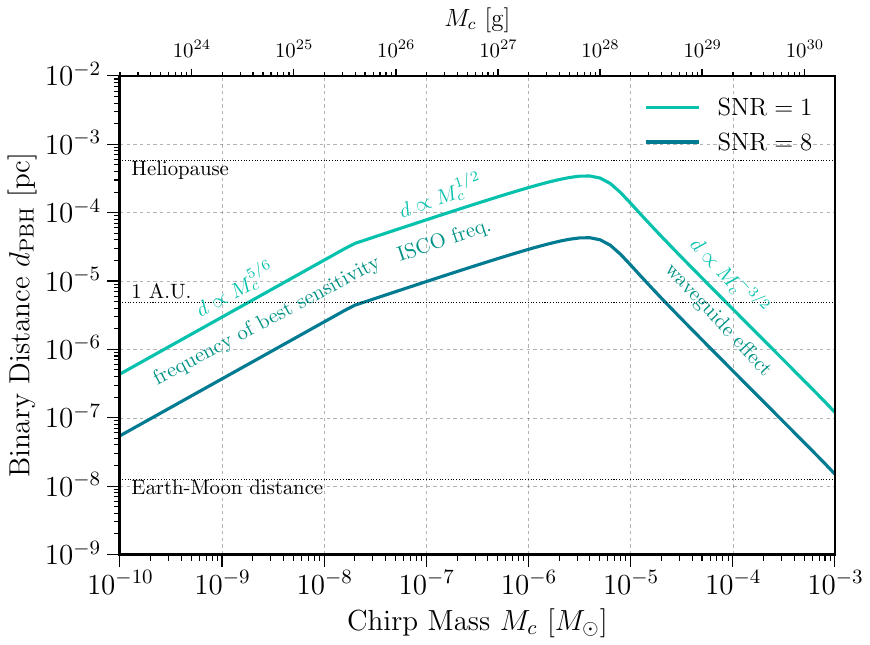}
    \caption{Maximum distance from the detector at which a merging PBH binary would be observable as a function of the binary chirp mass, assuming all PBHs have identical masses. The scaling of the sight distance as a function of the chirp mass is shown, as well as what dominates the various regions of the curve. The top line (light blue) assumes $S/N = 1$, while the lower line (dark blue) assumes $S/N = 8$. We fix the apparatus length $\ell = 2\,\text{m}$ and the detector area $\pi R^2 = 1\,\text{m}^2$, and assume a GW propagating along the detector axis ($\theta = 0$).
    }
    \label{fig:SharkFin}
\end{figure}

Assuming matched filtering for the coalescing binary signal, the SNR is given by~\cite{Maggiore:2007ulw}\footnote{Note the factor of 2 difference with respect to Ref.~\cite{Maggiore:2007ulw}'s expression, due to our use of two-sided PSDs.}
\begin{align}
    \text{SNR} \simeq\left( 2\int_0^\infty df\, \frac{|\tilde{h}(f)|^2}{S_h^{\rm noise}(f)}\right)^{1/2} \ ,
    \label{eq:snr}
\end{align}
which can be inverted to compute the maximum distance $d$ at which a binary merger can be seen,
\begin{align}
    d = \left(\frac{5}{6}\right)^{1/2} \frac{1}{\pi^{2/3}}  \left( G M_c \right)^{5/6} \left( \int_{f_{\rm min}}^{f_{\rm max}} df \, \frac{f^{-7/3} }{S_h^{\rm noise}(f)} \right)^{1/2} \frac{1}{\text{SNR}}\ .
    \label{eq:sightDistance}
\end{align}
Here, we have restricted the integral to the frequency range $[f_{\rm min}, f_{\rm max}]$ in which the detector and source system both have support. In practice, $f_{\rm max}$ can be set either by the maximum frequency of the binary inspiral or the maximum frequency the detector is capable of measuring. In \cref{fig:SharkFin}, we plot this distance reach for (primordial) black hole binary mergers as a function of the chirp mass, assuming the same detector parameters as in \cref{fig:sens} for the fully broadband MADMAX setup (no disks).

There are three qualitatively different regimes in \cref{fig:sens}. For small chirp masses, the integral in \cref{eq:sightDistance} is dominated by the frequency at which the detector is most sensitive, which for the parameters chosen is approximately $f_{\rm best} \sim \SI{e11}{Hz}$. The integral is then approximately
\begin{align}
     \int_{f_{\rm min}}^{f_{\rm max}} df \, \frac{f^{-7/3} }{S_h^{\rm noise}(f)} \sim \frac{f_{\rm best}^{-4/3}}{S_h^{\rm noise}(f_{\rm best})} \propto f_{\rm best}^{2/3} \,,
    \label{eq:SNR-integral}
\end{align}
with $S_h^\text{noise}(f)$ given in \cref{eq:Shnoise} in the main text.  Note that \cref{eq:SNR-integral} is independent of $M_c$, such that the scaling of $d$ with $M_c$ is given solely by the $M_c^{5/6}$ factor in \cref{eq:sightDistance}.

For chirp masses above $M_c \gtrsim 10^{-8}\,M_\odot$, we observe a different scaling, namely $d \propto M_c^{1/2}$. In this regime, the integral in \cref{eq:sightDistance} is dominated by the maximum frequency the binary reaches at the innermost stable circular orbit (ISCO), given by  $f_{\rm ISCO} \sim \SI{2e3}{Hz} \times (M_\odot/M_c)$. More precisely, the frequency integral is well approximated by
\begin{align}
     \int_{f_{\rm min}}^{f_{\rm max}} df \, \frac{f^{-7/3} }{S_h^{\rm noise}(f)} \sim \frac{f_{\rm ISCO}^{-4/3}}{S_h^{\rm noise}(f_{\rm ISCO})} \propto M_c^{-2/3} \ ,
\end{align}
%
in this regime, such that we obtain the observed chirp mass scaling in Fig.~\ref{fig:SharkFin}.

Finally, for chirp masses above $M_c \gtrsim 4\times 10^{-6}$, the sensitivity drops as $d \propto M_c^{-3/2}$. In this regime, the ISCO frequency is lower than the frequencies where the detector is most sensitive, and only reaches the frequencies where the waveguide effect is in force. The result is that rather than $S_h^{\rm noise} \propto 1/f^2$ as for smaller chirp masses, we are in the $S_h^{\rm noise} \propto 1/f^6$ regime. 
The frequency integral in \cref{eq:sightDistance}
becomes
\begin{align}
     \int_{f_{\rm min}}^{f_{\rm max}} df \, \frac{f^{-7/3} }{S_h^{\rm noise}(f)}  \propto f_{\rm ISCO}^{14/3} \propto M_c^{-14/3} \ ,
\end{align}
leading to the $d \propto M_c^{-3/2}$ scaling we observe in \cref{fig:SharkFin} for $M_c \lesssim 10^{-4}$. 

The distance reach shown in \cref{fig:SharkFin} needs to be seen in the context of the expected rate of PBH merger events. A rough estimate of this rate is given by~\cite{Raidal:2018bbj, Hutsi:2020sol, Franciolini:2021nvv}
\begin{align}
    R_0 \simeq \SI{e-2}{kpc^{-3} \, yr^{-1}} \, \times 
               f_\text{PBH}^{53/37}
               \left( \frac{\delta}{2 \cdot 10^5} \right)
               \left(\frac{M_c}{10^{-5} M_\odot} \right)^{-32/37} \,,
\end{align}
where $f_\text{PBH} \leq 1$ is the fraction of dark matter that is in the form of primordial black holes, and $\delta$ is the local dark matter overdensity in the Milky Way halo compared to the average cosmological dark matter density. Taking into account microlensing constraints which limit $f_\text{PBH} \lesssim 0.01$~\cite{Carr:2021bzv}, this indicates that a reach of at least tens of kpc is needed to achieve reasonable rate of observed events.

When computing the expected event rate for a given detector sensitivity more accurately, we need to take into account also the field of view of the detector. This amounts to considering the angular average of $S_h^\text{sig}$, where for simplicity we conservatively only take into account GWs passing through all the disks, such that the angular averaged $S_h^\text{sig}$ becomes
\begin{align}
\langle S_h^\text{sig} \rangle_\theta= \frac{1}{2 \pi} \int_{- \arctan(2R/\ell)}^{\arctan(2R/\ell)} \cos \theta \, d\theta \; S_h^\text{sig} (\theta = 0) = \frac{2 R}{\pi \ell \sqrt{1 + \frac{4 R^2}{\ell^2}}} \, S_h^\text{sig} (\theta = 0) \equiv \eta \, S_h^\text{sig} (\theta = 0)\,.
\end{align}
From the definition of the strain-equivalent noise PSD in \cref{eq:Shnoise}, we see that this amounts to replacing $S_h^\text{noise} \mapsto S_h^\text{noise} / \eta $ in \cref{eq:snr}.
As in the main text, we have here restricted ourselves to GWs in the $x$--$z$ plane for simplicity. For detectors with $R \sim \ell$ the factor $\eta$ is order one, however for $R \ll \ell$ it will be much smaller, significantly limiting the effective sensitivity. In particular, for the parameters of \cref{fig:SharkFin}, $\eta =  1/(\sqrt{2}\pi ) $.

\bibliographystyle{utphys}
\bibliography{refs}

\end{document}